\documentclass[letterpaper,11pt]{article}
\usepackage{natbib}
\usepackage{tabularx} 
\usepackage{amsmath}  
\usepackage{graphicx} 
\usepackage[margin=1in,letterpaper]{geometry} 
\usepackage[final]{hyperref} 
\usepackage[dvipsnames]{xcolor}
\usepackage{amssymb}
\usepackage{booktabs}
\usepackage{longtable}
\usepackage{array}
\usepackage{multirow}
\usepackage{wrapfig}
\usepackage{float}
\usepackage{colortbl}
\usepackage{pdflscape}
\usepackage{tabu}
\usepackage{threeparttable}
\usepackage{threeparttablex}
\usepackage[normalem]{ulem}
\usepackage{makecell}
\usepackage{multirow}
\usepackage{longtable}
\usepackage{subfig}

\usepackage{arttab}

\hypersetup{colorlinks,allcolors=black}

\providecommand{\keywords}[1]
{
  \small	
  \textbf{\textit{Keywords: }} #1
}

\begin{document}

\title{\textbf{Construction and Hedging of Equity Index Options Portfolios}}

\author{Maciej Wysocki$^{1}$\thanks{Corresponding author: m.wysocki9@uw.edu.pl, 44/50 Dluga, Warsaw, Poland. ORCID: 0000-0002-1693-1438} \thanks{This research was supported by IDUB University of Warsaw (grant number 501-D124-20-0004410)}, Robert Ślepaczuk$^{1}$\thanks{ORCID: 0000-0001-5227-2014} \\ \\
\small $^{1}$Department of Quantitative Finance and Machine Learning \\ 
\small Faculty of Economic Sciences, University of Warsaw \\
\small Quantitative Finance Research Group}
\date{}

\maketitle

\renewcommand*\abstractname{Abstract}
\begin{abstract}
This research presents a comprehensive evaluation of systematic index option-writing strategies, focusing on S\&P500 index options. We compare the performance of hedging strategies using the Black-Scholes-Merton (BSM) model and the Variance-Gamma (VG) model, emphasizing varying moneyness levels and different sizing methods based on delta and the VIX Index. The study employs 1-minute data of S\&P500 index options and index quotes spanning from 2018 to 2023. The analysis benchmarks hedged strategies against buy-and-hold and naked option-writing strategies, with a focus on risk-adjusted performance metrics including transaction costs. Portfolio delta approximations are derived using implied volatility for the BSM model and market-calibrated parameters for the VG model. Key findings reveal that systematic option-writing strategies can potentially yield superior returns compared to buy-and-hold benchmarks. The BSM model generally provided better hedging outcomes than the VG model, although the VG model showed profitability in certain naked strategies as a tool for position sizing. In terms of rehedging frequency, we found that intraday heding in 130-minute intervals provided both reliable protection against adverse market movements and a satisfactory returns profile.

\keywords{S\&P500 Index options, Option Pricing Models, Black-Scholes-Merton model, Variance-Gamma model, Implied Volatility, Volatility Risk Premium, Volatility Spreads, Dynamic Hedging}

{\textit{\textbf{JEL codes: }}C4, C14, C45, C53, C58, G13}
\end{abstract}

\vspace{10pt}


\section{Introduction}
The primary focus of this study encompasses two distinct yet interrelated facets: the theoretical aspects of the models under examination and their practical implementation for options trading. The significance of this investigation is supported by the growing scale of options portfolios within financial institutions, constructed through complex investment strategies and market-making processes. The consequential impact of these institutions on the market is well-documented (\cite{el_kalak_reviewing_2016}), necessitating a thorough understanding of the theoretical aspects and practical implications of options trading models. Therefore, adequate risk assessment and portfolio hedging became crucial elements of investment activities, especially during periods of rapid volatility fluctuations. Therefore, the efficacy of hedging strategies is evaluated across both low and high-market volatility regimes. 


To empirically validate different approaches, algorithmic investment strategies were employed for trading in index options and a related exchange-traded fund (ETF). Based on the concept of volatility risk premium (VRP) and aiming to exploit options premiums by selling volatility, systematic option writing strategies, such as short calls, short puts, and volatility spreads, were utilized with various hedging schemes and sizing methodologies. The models used include the well-established BSM model (\cite{black_pricing_1973}, \cite{merton_theory_1973}), incorporating options' implied volatility as its volatility estimator, and the VG model (\cite{madan_variance_1998}).

The choice of these methodologies is based both on a comprehensive synthesis of existing literature and a pragmatic standpoint. Volatility spreads, including short straddles and short strangles, represent practically employed strategies within the investor community and have been the subject of extensive academic research (\cite{gao_anticipating_2017}, \cite{hong_profitability_2018}, \cite{s_p_choosing_2022}). The BSM model is well-established among market practitioners for its simplicity and reliability (\cite{karagozoglu_option_2022}). On the other hand, it is known for its certain limitations, such as volatility smiles and skewness premiums (\cite{corrado_skewness_1996}). Over the years, research efforts have yielded alternative approaches, such as the Variance-Gamma (VG) model incorporating the variance-gamma stochastic process. This selection of models combines theoretical grounding with practical applicability, offering an exploration of the complexities inherent in options trading methodologies.


While there is a rich literature on options pricing, hedging, and trading, this research stands as a heavily empirical contribution. Although the performance of the BSM and the VG models in options pricing is present in the literature (see e.g., \cite{lam_empirical_2002}, \cite{daal_empirical_2005}), most articles focus on the comparison of models' pricing capabilities rather than their hedging performance. Some studies have compared hedging methodologies (\cite{han_robust_2015}), but the focus is often on single options hedged once a day. Empirical analyses of volatility spreads and other option writing strategies are widely available in the literature (see e.g., \cite{chaput_option_2003}, \cite{hill_finding_2006}, \cite{nikolopoulos_options_2007}, \cite{black_35-year_2022}). 

There are numerous studies analyzing dynamic delta hedging of options on various asset classes, e.g. cryptocurrency options (\cite{alexander2023delta}), equity indices (\cite{XIA2023106898}, \cite{MOZUMDER2016285},\cite{doi:10.1080/14697688.2022.2136037}), currencies (\cite{ANDEREGG2022102627}), singles stocks (\cite{HULL2017180}), commodities (\cite{HULL2017180}), interest rates (\cite{HULL2017180}), and based on simulated data (\cite{10.1145/3383455.3422532}). However, many of these papers lack realism in market conditions and focus on the performance of derivative instruments. This study aims to bridge the gap by comparing two popular option pricing models in the practical setting of systematic option-writing strategies using intraday hedging and high-frequency data.

The contribution of this study is twofold. Firstly, we provided empirical evidence for choosing the right hedging specification for systematic option-writing strategies on one of the most liquid derivatives markets globally. Secondly, it conducts an extensive backtest of option-writing strategies, introducing two sizing methods: leverage-aware delta-based sizing and a novel volatility spikes-avoiding VIX-rank methodology. While results suggest that each strategy requires individual analysis, the overarching findings indicate that the BSM model offers more accurate hedging than the VG model when considering portfolio CVaR and VaR. Strategies using the Variance-Gamma model may be more profitable but also riskier. The VIX-rank methodology proves stable, providing reliable returns, although generally underperforming benchmarks in plain annualized returns. Regarding hedging, intraday adjustments every 130 minutes yield the best balance between benefits and costs.

The paper is organized as follows. Section 2 describes the methodology and backtesting procedures, Section 3 describes the data, Section 4 presents the results, and Section 5 concludes. 

\section{Methodology}
\subsection{Black-Scholes-Merton Model}
The BSM model (\cite{black_pricing_1973}, \cite{merton_theory_1973}) was introduced as a comprehensive model for pricing European options with closed-form solutions based on partial differential equations (PDEs). The main PDE underlying the model is:
\begin{equation}
\frac{\partial V}{\partial t} + \frac{1}{2}\sigma^2S^2\frac{\partial^2 V}{\partial S^2} + rS\frac{\partial V}{\partial S} - rV = 0
\label{eq:1}
\end{equation}
where $V$ is the price of the option at time $t$, $S$ is the price of the underlying at time $t$, $r$ is the risk-free interest rate and $\sigma$ is the volatility of the underlying asset. The equation \ref{eq:1} indicates that it is possible to neutralize the delta risk by buying or selling the underlying asset with money borrowed at the risk-free rate. The prices of European options can be calculated with the following formulas (\cite{black_pricing_1973}, \cite{merton_theory_1973}):
\begin{equation}
C(S, t) = S_te^{-qt} \cdot \Phi(d_1) - Ke^{-rt} \cdot \Phi(d_2)
\label{eq:bscall}
\end{equation}

\begin{equation}
P(S, t) = Ke^{-rt} \cdot \Phi(-d_2) - S_te^{-qt} \cdot \Phi(-d_1)
\label{eq:bsput}
\end{equation}

\begin{equation}
d_1 = \frac{\ln(\frac{S_t}{K}) + (r - q)t}{\sigma\sqrt{t}} + \frac{\sigma\sqrt{t}}{2}
\label{eq:bsd1}
\end{equation}

\begin{equation}
d_2 = \frac{\ln(\frac{S_t}{K}) + (r - q)t}{\sigma\sqrt{t}} - \frac{\sigma\sqrt{t}}{2} = d_1 - \sigma\sqrt{t}
\label{eq:bsd2}
\end{equation}
where $C$ is the price of the call option, $P$ is the price of the put option, $q$ denotes the continuous dividend rate, $K$ stands for the strike price and $\Phi(\cdot)$ represents the cumulative distribution function of the standard normal distribution. The only unknown parameter that requires estimation is $\sigma$, which denotes the underlying asset's volatility.

As options prices can be obtained from the market, and other parameters are known, the BSM model can be adapted to produce volatility instead of prices. In this study, to determine the delta value, we utilized the last available implied volatility for that specific option. Therefore, at each timestamp $t$, when delta was required, we utilized $\sigma^{IV}_{i, t-1}$, which denotes the $i$-th option's BSM implied volatility from timestamp $t-1$, essentially from the previous minute.

The options' deltas were calculated with formulas \ref{eq:bscalldelta} and \ref{eq:bsputdelta}.

\begin{equation}
\Delta_C = \Phi(d_1)
\label{eq:bscalldelta}
\end{equation}

\begin{equation}
\Delta_P = \Phi(d_1) - 1
\label{eq:bsputdelta}
\end{equation}

\subsection{Variance-Gamma Model}
The VG model is based on a generalization of the Brownian motion, used to describe the process of asset price. Equation \ref{eq:vgprocess} presents the Variance-Gamma process as a Brownian motion $W(t)$ with drift $\theta t$ and the time change modeled with a gamma process $\Gamma(t, 1, \nu)$.
\begin{equation}
X^{VG}(t; \sigma; \nu; \theta) = \theta \Gamma(t; 1; \nu) + \sigma W(\Gamma(t; 1; \nu))
\label{eq:vgprocess}
\end{equation}

Under the VG model, the evolution of an asset price is described with equation \ref{eq:vgassetprice}. 

\begin{equation}
S_t = S_0e^{(r-q+\omega)t + L(t)}, t \geq 0
\label{eq:vgassetprice}
\end{equation}
where $r$ is the rate of return on the asset, $q$ is the dividend yield, $L(t)$ is a L\`evy process, and $\omega = -\psi_X(-i)$. The characteristic function of $L(t)$ satisfies the following condition:
\begin{equation}
\psi_{X(t)}(\xi):=\mathbb{E}[e^{iX(t)\xi}]=e^{t\psi_L(\xi)}
\label{eq:vgcharacteristic}
\end{equation}
where $\psi_L(\xi)$ is the L\`evy symbol of process $X(t)$. VG symbol is expressed as:

\begin{equation}
\psi_{X}(\xi) = -\frac{\sigma^2}{2} - \frac{1}{\nu}\ln{(1-i\nu\theta\xi + \nu\xi^2\frac{\sigma^2}{2})}
\label{eq:symbol}
\end{equation}
In this study, we will use a convexity-corrected symbol calculated as:
\begin{equation}
\psi_{X}^{C}(\xi) = i(\nu-\psi_{X}(-i))\xi+\psi_{X}(\xi)
\label{eq:convexitysymbol}
\end{equation}
which ensures that $\mathbb{E}[e^{X(t)}] = 1$ for any $\phi_L(\xi)$.

To price options under the L\`evy processes, we employed the frame projection with the Fast Fourier Transform (FFT) methodology (\cite{kirkby_efficient_2015}), facilitating quick and efficient pricing. This same methodology was employed for model calibration, i.e., estimating its parameters based on market prices. Calibration was performed every 30 minutes during each trading session to ensure the parameters remained current. A least squares optimizer was used to determine $\{\sigma, \nu, \theta\}$ parameters accurately reflecting prevailing market conditions. To calculate deltas, we used a finite differences approximation as described in equation \ref{eq:numericaldelta}.

To calculate deltas, we employed a finite differences approximation as described in Equation \ref{eq:numericaldelta}.

\begin{equation}
\Delta(t) = \frac{V(t, S + dS) - V(t, S - dS)}{2dS}
\label{eq:numericaldelta}
\end{equation}

\subsection{Backtesting}
We invested considerable effort in realistically backtesting all the strategies to provide valuable insights into real market conditions. Our commissions model reflects fees from the Interactive Brokers brokerage firm\footnote{See: \href{https://www.interactivebrokers.com/en/pricing/commissions-home.php}{Interactive Brokers Commissions} for more information.}. Furthermore, we adopted a stringent assumption regarding order fills, considering that buy and sell orders are executed at the midpoint of the bid and ask prices, covering half of the bid-ask spread.

We employed additional methodologies to benchmark and compare the results of different model-based strategies. First, the buy-and-hold strategy assumed that at the beginning of the backtesting period, the strategy invested all available cash into the S\&P500 Index and held it until the end of the backtesting period. Additionally, we presented results of \textit{naked} strategies, which were the same option-selling strategies with no hedging at all. These benchmarks enabled a realistic assessment of the performance of model-based strategies and facilitated a comparison with what the market offered. 

\subsection{Performance Metrics}
To assess strategies' performances, we utilized several performance metrics considering both the profitability and riskiness of the investment systems. The following definitions were used:

\begin{itemize}
    \item Daily Returns
\begin{equation}
r_i:=\frac{p_i - p_{i-1}}{p_t}
\label{eq:dailyret}
\end{equation}
Where  $p_t$ is asset price on day $i$ of backtest.
    \item Annualized Compounded Rate of Returns
\begin{equation}
ARC:=\prod_{i=1}^{n}(1+r_i^{\frac{252}{n}}) - 1
\label{eq:arc}
\end{equation}
Where $n$ is the number of days in the whole backtesting period.
    \item Annualized Standard Deviation
\begin{equation}
aSD:=\sqrt{252} \cdot \sqrt{\frac{1}{n-1}\sum_{i=1}^{n}(r_i - \overline{r})^2}
\label{eq:asd}
\end{equation}
Where $\overline{r}$ is the average of daily returns. 
    \item Maximum Drawdown
\begin{equation}
MD:=\sup_{(x, y)\in\big\{(t_i, t_j) \in \mathbb{R}: t_i < t_j\big\}}\frac{p_x - p_y}{p_x}
\label{eq:md}
\end{equation}
Where $p$ is the total equity value.
    \item Maximum Loss Duration
\begin{equation}
MLD:=\frac{y-x}{252}
\label{eq:mld}
\end{equation}
Where $x, y$ are days defined in formula \ref{eq:md} and indicate days of consecutive local maxima of the equity values.
    \item Information Ratios
\begin{equation}
IR:=\frac{aRC}{aSD}
\label{eq:ir1}
\end{equation}
\begin{equation}
IR^{**}:=\frac{IR \cdot sign(aRC) \cdot aRC}{MD}
\label{eq:ir2}
\end{equation}
\begin{equation}
IR^{***}:=\frac{aRC^3}{aSD \cdot MD \cdot MLD} * 1000
\label{eq:IR***}
\end{equation}
    \item Value-at-Risk
\begin{equation}
VaR_{\alpha}(X) := -\inf\big\{x \in \mathbb{R}: F_{X}(x) \geq \alpha \big\}
\label{eq:var}
\end{equation}
We used the historical VaR of portfolio returns at a 95\% confidence level, which means that we used the 5th percentile of portfolio returns to estimate VaR.
    \item Conditional Value-at-Risk
\begin{equation}
CVaR_{\alpha}{X}:=-\frac{1}{\alpha}\int_{0}^{\alpha} VaR(\gamma)d\gamma
\label{eq:cvar}
\end{equation}
We estimated CVaR (expected shortfall, ES) as the average of returns below $VaR_{\alpha}(X)$.
\end{itemize}

In each case, we assumed the number of trading days in a year is 252.

\subsection{Investment Strategies}
We employed four strategies related to selling options and the volatility risk premium concept. The strategies involve selling options and hedging positions against the exposure to changes in the underlying asset's value. The assumptions were held the same for all strategies and they are described in Table \ref{tab:assumptions}.

\begin{table}[!ht]
\caption{The assumptions of tested strategies.}
\label{tab:assumptions}
\resizebox{\columnwidth}{!}{%
\begin{tabular}{@{}cccc@{}}
\toprule
Step &
  Name &
  Description \\ \midrule \addlinespace[0.75em]
1 &
  strategy open &
  Selling options with 7 days to expiration (7DTE) and holding them until expiry \\ \addlinespace[0.75em]
2 &
  strategy reopen &
  \begin{tabular}[c]{@{}c@{}}At each expiration date, before the market close and options settlement,\\ we opened another position according to the given strategy so that the system is always in the market.\end{tabular}  \\ \addlinespace[0.75em]
3 &
  frequency of hedging &
  \begin{tabular}[c]{@{}c@{}}In the case of hedged strategies, the options positions were hedged every {30 minutes, 130 minutes, 1 day}.\end{tabular} \\ \addlinespace[0.75em]
4 &
  time of hedging &
  \begin{tabular}[c]{@{}c@{}}The daily hedging position adjustment was done 30 minutes before the NYSE market session closed.\end{tabular} \\ \addlinespace[0.75em]
5 &
 \begin{tabular}[c]{@{}c@{}}basis instrument used \\ for hedging purposes\end{tabular} &
  \begin{tabular}[c]{@{}c@{}} The delta hedging procedure involves buying and selling the ETF tracking the underlying index. \end{tabular} \\ \addlinespace[0.75em]
6 &
  details of hedging &
  \begin{tabular}[c]{@{}c@{}}The quantity of ETF shares was determined so that the net delta of the total portfolio was equal to zero. \end{tabular} \\ \addlinespace[0.75em]
7 &
  benchmark &
  \begin{tabular}[c]{@{}c@{}}The results for naked (unhedged) strategies are presented \\ to serve as a benchmark for assessing hedged strategies’ performance.\end{tabular}  \\
\bottomrule
\end{tabular}%
}
\source{The procedure described in this table is repeated throughout the whole data set for all four strategies tested in this research.}
\end{table}

Table \ref{tab:strategies} describes options strategies considered in the trading system.

\begin{table}[!ht]
\caption{Option strategies definitions.}
\label{tab:strategies}
\resizebox{\columnwidth}{!}{%
\begin{tabular}{@{}ccccc@{}}
\toprule
Name &
  Description &
  Profit &
  Loss &
  Position sizing \\ \midrule \addlinespace[0.75em]
Short Call &
  \begin{tabular}[c]{@{}c@{}}short at-the-money (ATM) call or\\ out-of-the-money (OTM) call\end{tabular} &
  limited = total premium &
  unlimited & 
  \begin{tabular}[c]{@{}c@{}}adjusted dynamically\\ based on delta or VIX rank\end{tabular} \\ \addlinespace[0.75em]
Short Put &
    \begin{tabular}[c]{@{}c@{}}short at-the-money (ATM) call or\\ out-of-the-money (OTM) put\end{tabular} &
  limited = total premium &
  unlimited & 
  \begin{tabular}[c]{@{}c@{}}adjusted dynamically\\ based on delta or VIX rank\end{tabular} \\ \addlinespace[0.75em]
Short Straddle &
  \begin{tabular}[c]{@{}c@{}}short at-the-money (ATM) call and\\ short at-the-money (ATM) put\\ with the same strike and maturity\end{tabular} &
  limited = total premium &
  unlimited & 
  \begin{tabular}[c]{@{}c@{}}adjusted dynamically\\ based on delta or VIX rank\end{tabular} \\ \addlinespace[0.75em]
Short Strangle &
  \begin{tabular}[c]{@{}c@{}}short out-of-the-money (OTM) call and\\ out-of-the-money (OTM) put\\ with the same maturity, but different strikes\end{tabular} &
  limited = total premium &
  unlimited & 
  \begin{tabular}[c]{@{}c@{}}adjusted dynamically\\ based on delta or VIX rank\end{tabular} \\ 
\bottomrule
\end{tabular}%
}
\source{In practice, straddle and strangle reflect the same idea and straddle is a special case of strangle, when both options have the same strike price.}
\end{table}

\subsubsection{Delta-based sizing}
The delta-based sizing was designed to hold the portfolio's leverage at a nearly constant level throughout the investing period. To arrive at a position size for time $t$ we used formula \ref{eq:2}. 

\begin{equation}
Q_t = \Bigg\lfloor{\frac{PV_t}{\sum_{i} K_i \cdot |\Delta_{i, t}| \cdot M}}\Bigg\rfloor
\label{eq:2}
\end{equation}
$Q_t$ stands for the number of contracts, $K_i$ is the $i$-th option's strike price, $\Delta_{i, t}$ is the option's delta at time $t$ and $M$ is the contract multiplier, which is equal to 100. $PV_t$ is the portfolio value at time $t$. This approach incorporates the position's total delta and its leverage compared to the actual portfolio value. Investing with a highly leveraged portfolio of short options is related to bearing a tremendous level of risk and increasing margin requirements, which investors should intentionally avoid. 

\subsubsection{VIX-based sizing}
The CBOE Volatility Index (VIX) is a measure of the 30-day expected volatility of the US stock market based on prices of the S\&P500 Index and the corresponding options on that index. Since this study addresses investment strategies involving that index and its options, using the industry-recognized volatility index was the natural direction. The VIX-based sizing was designed to avoid holding large positions during volatile periods. To arrive at a position size for time $t$ we used formula \ref{eq:vix-sizing}. 

\begin{equation}
Q_t = \Bigg\lfloor{\frac{PV_t}{SPX_t} \cdot \rho \cdot (1 - P_{rank}(VIX_t))}\Bigg\rfloor
\label{eq:vix-sizing}
\end{equation}
$Q_t$ stands for the number of contracts, $PV_t$ is the portfolio value at time $t$, $\rho$ is a risk factor, $P_{rank}(VIX_t)$ is a percentile rank of the VIX index at time $t$ compared to the last 252 trading days (1 trading year). $\rho$ is subject to optimizations and its choice is up to investor preferences, as higher values of this parameter imply greater risk, but also potential for higher returns. The introduction of VIX's percentile rank in such a setting intends to take larger positions in relatively low-volatility periods and smaller positions in relatively high-volatility periods.  

\section{Data}
This study is based on 1-minute quotes of the S\&P500 Index and the corresponding quotes of S\&P500 European index options from the Chicago Board Options Exchange (CBOE). Specifically, the strategies were employed to trade the SPXW options, sometimes referred to as \textit{weeklies}. Those products are European-style options settled with cash and exercised at expiration. The slight difference from the standard SPX options is that the SPXW are P.M. settled, meaning that they are tradable until the end of the session on the expiration date. The contract multiplier is equal to 100. For each minute, we utilized bar data (open, high, low, close) as well as bid and ask quotes.

In addition to the market quotes of the S\&P500 Index and its options, we relied on daily quotes of the VIX index from CBOE. The proxy for the risk-free rate was the U.S. 3-Month Treasury Bill rate, and for the BSM model, we utilized the S\&P500 Index continuous dividend yield.

\begin{table}[!h]
\caption{Summary statistics of daily close returns.}
\label{tab:summary}
\resizebox{\columnwidth}{!}{%
\begin{tabular}{@{}ccccccccccccc@{}}
\toprule
Mean    & St.D.   & Var     & Min     & P10     & P25     & P50    & P75    & P90    & Max    & Skew   & Kurtosis & S-W Test \\ \midrule
0,00046 & 0,01298 & 0,00017 & -0,1202 & -0,0128 & -0,0049 & 0,0009 & 0,0069 & 0,0135 & 0,0942 & -0,509 & 12,878   & 0,881***    \\ \bottomrule
\end{tabular}%
}
\source{The returns were calculated for a daily close average of the bid and ask prices. P10, P25, P50, P75, P90 stand for 10th, 25th, 50th, 75th and 90th percentiles respectively. The S-W Test column contains test statistics for the Shapiro-Wilk normality test. *, **, *** indicate statistical significance on 0.1, 0.05, and 0.01 significance levels respectively.}
\end{table}

Table \ref{tab:summary} presets the basic summary statistics of the S\&P500 Index daily returns, where the price is represented as the average of bid and ask prices. The average daily return is 0.046\%, indicating a slight positive drift over the analyzed period. The standard deviation indicates that the returns can deviate by approximately 1.3\% from the mean on a typical day, highlighting a moderate level of market volatility. The minimum return of -0.1202 represents the largest single-day loss in the dataset, which occurred during the COVID-19 crash in March 2020. The percentile values indicate a skewed distribution with more frequent small losses and gains, with the potential for significant adverse movements. The maximum return of around 9.4\% was also noted in March 2020, when the COVID-19 crash turned into a market rally. The negative skewness indicates that the distribution of returns has a longer left tail, suggesting that large negative returns are more frequent than large positive returns. The kurtosis value indicates a leptokurtic distribution, meaning the returns have fat tails and a sharp peak compared to a normal distribution. This implies a higher probability of extreme returns, both positive and negative, than what would be expected in a normal distribution. Finally, the Shapiro-Wilk test statistic of 0.881, states that the daily returns are not normally distributed at the 1\% significance level. This non-normality is also supported by the high kurtosis and skewness values.

Overall, the summary statistics reveal that the S\&P500 Index daily returns exhibit slight positive mean returns, moderate volatility, and a distribution characterized by negative skewness and leptokurtosis. The non-normality of the returns distribution is evident, indicating the presence of fat tails and extreme values, which are also crucial considerations for risk management and option-writing strategies.

\begin{figure}[h]
\caption{\label{fig:indexvalue} S\&P500 Index quotes from 02-01-2018 to 29-12-2023.}
\begin{center}
\includegraphics[width=1\columnwidth]{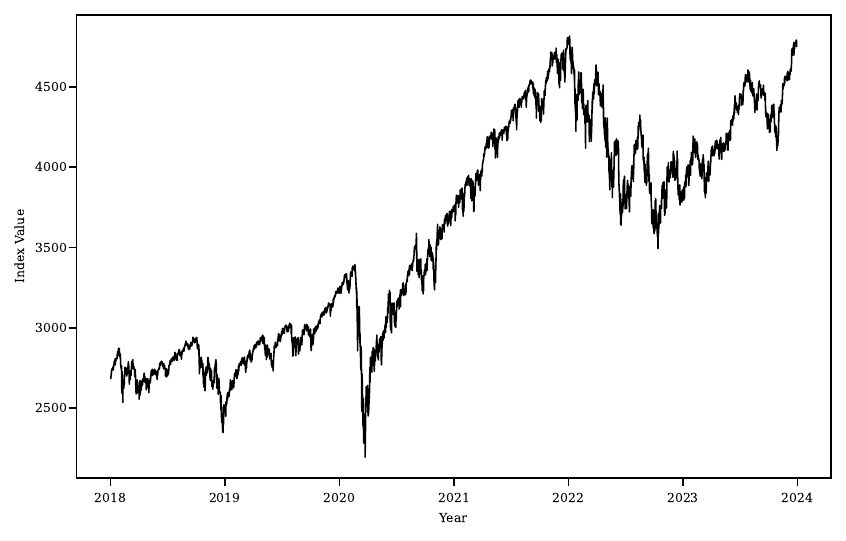}
\end{center}
\source{The values presented on the graph are average of 1-minute bid-ask quotes}
\end{figure}

Our investment strategy backtesting started at the beginning of 2018 and lasted until the very end of 2023. Throughout this period, there were various market conditions, including instances of exceptionally high volatility (e.g., the COVID-19 pandemic outbreak), rapid growth (e.g., in 2021), and stabilization (e.g., in 2022), which can all be clearly observed on Figure \ref{fig:indexvalue}. These diverse economic circumstances facilitated a comprehensive examination of the investment strategies and enabled us to conclude their performance in different environments.

\section{Results}

\subsection{Short Call}
Table \ref{tab:short-call-results} presents the results of the short-call strategies for each backtested combination. Regarding PnL's standard deviation and drawdown, strategies with VIX-based sizing performed well, delivering stable results with minimal drawdowns. The 10\% OTM short call with VIX-based sizing produced the best results.

When comparing the BSM-hedged strategies with delta sizing at different hedging intervals (30 minutes, 130 minutes, and once a day) to the naked short call strategies, the naked strategies exhibit significant variability in returns, with high standard deviations (aSD) and maximum drawdowns (MD), reflecting higher risk exposure. For instance, the naked short call with 0\% OTM using delta sizing under BSM has a negative annualized return (aRC) of -9.769\% and a standard deviation of 0.204. This high risk is mitigated by the hedged strategies. The BSM-hedged strategy hedged every 30 minutes at 0\% OTM shows an aRC of 0.878\% and a lower aSD of 0.1, indicating a substantial reduction in risk and an improvement in return consistency. Similarly, hedging a strategy with delta sizing every 130 minutes at 0\% OTM yields a higher aRC of 3.347\% with an aSD of 0.113, further highlighting the benefits of more frequent rehedging. The once-daily rehedging approach also improves performance metrics compared to the naked approach but not as significantly as the more frequent intervals. Notably, rehedging every 130 minutes at 2\% OTM produces the highest aRC of 6.504\% with a relatively low aSD of 0.102, demonstrating that optimal rehedging intervals and OTM levels can significantly enhance strategy performance. These hedged strategies also show improved information ratios, indicating better risk-adjusted returns. The BSM strategy rehedging every 130 minutes at 5\% OTM achieves an IR*** of 13.962, compared to generally lower IRs in naked strategies.

Comparing strategies with VIX sizing and BSM hedging at different rehedging intervals to VIX sizing naked strategies also reveals notable improvements in performance and risk management. The naked VIX sizing strategies, particularly at 0\% OTM, show negative annualized returns and relatively high standard deviations. For instance, the naked VIX sizing strategy at 0\% OTM records an aRC of -3.358\%, an aSD of 0.062, and an MD of 0.195. In contrast, the VIX sizing with BSM hedging significantly enhances these metrics. The strategy rehedging every 30 minutes at 0\% OTM yields an aRC of -0.333\% and an aSD of 0.031, reducing both losses and volatility compared to the naked approach. Similarly, rehedging every 130 minutes at 0\% OTM results in an aRC of 0.793\% and an aSD of 0.036, demonstrating improved returns. The once-daily rehedging strategy also outperforms its naked counterpart, with an aRC of 1.106\% and an aSD of 0.045. These hedged strategies also show superior information ratios, indicating better risk-adjusted returns. Furthermore, the CVaR and VaR metrics for hedged strategies are generally lower, indicating reduced potential for extreme losses. For example, the BSM-hedged strategy rehedging every 30 minutes at 5\% OTM records a CVaR of 0.303\% and a VaR of -0.008\%, substantially better than the naked equivalent. Overall, the combination of VIX sizing with BSM hedging at optimal rehedging intervals provides more robust performance, enhancing returns and minimizing risk compared to the naked strategies.

Regarding the Variance-Gamma model, the naked VG delta sizing strategy at 0\% OTM shows significant negative returns, elevated standard deviation, and a large maximum drawdown. The remaining naked strategies with this sizing methodology did not stay underwater but are also relatively volatile. Introducing VG hedging at different intervals markedly improves these metrics. The VG delta sizing strategy with 30-minute rehedging at 0\% OTM achieves an aRC of 3.721\%, an aSD of 0.109, and an MD of 0.156, indicating not only a shift to positive returns but also a reduction in volatility and drawdowns. Similarly, the strategy with 130-minute rehedging at 0\% OTM shows an aRC of -3.426\%, an aSD of 0.122, and an MD of 0.348, which is negative but better than the naked strategy. The once-daily rehedging strategy at 0\% OTM also improves upon the naked approach, with an aRC of -2.156\%, an aSD of 0.127, and an MD of 0.356. For the ATM short call, only the most frequent hedging was appropriate. The hedged strategies generally exhibit lower CVaR and VaR metrics, indicating reduced potential for extreme losses. Overall, incorporating VG hedging into delta sizing strategies, especially with frequent rehedging, markedly enhances returns and reduces risk compared to naked VG delta sizing strategies, showcasing the effectiveness of hedging in managing downside risk and improving overall strategy performance.

When comparing hedged strategies with delta sizing, it is not entirely clear which model is a winner. Regarding IR***, in 6 out of 12 cases, the BSM model provided better results, and in the remaining cases, the VG model outperformed. On the other hand, regarding CVaR, the BSM hedging resulted in outcomes closer to 0 in 11 out of 12 cases respectively. Similar conclusions can be drawn when analyzing the VaR measures, which indicate that for intraday hedging, the BSM model is a better choice, providing better value-at-risk measures. For strategies with VIX sizing, it is a very similar story, as in 6 out of 12 cases, the BSM-hedged strategies achieved higher IR***, and in the remaining cases, the VG-hedged strategies did. Generally, the BSM hedging also resulted in better VaR and CVaR metrics, indicating reduced portfolio riskiness.

In summary, the BSM-hedged strategies, whether applied to delta sizing or VIX sizing, tend to outperform the VG-hedged ones in terms of risk measures and risk-adjusted performance. In most cases, the BSM-based strategies outperformed their naked counterparts in terms of VaR and CVaR, indicating that hedging indeed contributed to protection against losses. Although VG-based strategies often yielded higher aRC than the BSM-based ones and higher IR*** than the naked strategies, they also tended to carry higher risk. Strategies with VIX-based sizing performed very similarly across both models used for hedging and experienced disturbances for 5\% and 10\% OTM calls. In terms of aRC and IR***, the VG model generally provided better results for VIX-based sizing. For almost all strategies, increasing the hedging frequency from once a day to every 130 minutes or from every 130 minutes to every 30 minutes improves the CVaR and VaR metrics. In terms of IR***, however, it is not that obvious and requires a case-by-case analysis.

\begin{table}[!ht]
\caption{Short call strategies results.}
\label{tab:short-call-results}
\resizebox{\columnwidth}{!}{%
\begin{tabular}{@{}cccccccccccccc@{}}
\toprule
\multicolumn{5}{c}{STRATEGY} &
  \multirow{2}{*}{aRC (\%)} &
  \multirow{2}{*}{aSD} &
  \multirow{2}{*}{MD} &
  \multirow{2}{*}{MLD} &
  \multirow{2}{*}{IR} &
  \multirow{2}{*}{IR2} &
  \multirow{2}{*}{IR3} &
  \multirow{2}{*}{CVaR (\%)} &
  \multirow{2}{*}{VaR (\%)} \\ \cmidrule(r){1-5}
OPTIONS    & MODEL & SIZING    & REHEDGING & \%OTM &                &       &       &       &        &        &        &        &                 \\ \midrule
-          & B\&H  & -         & -         & -     & \textbf{9,889} & 0,206 & 0,340 & 1,980 & 0,479  & 0,140  & 6,970  & -1,935 & -3,160          \\ \midrule
SHORT CALL & NAKED & DELTA BSM & -         & 0\%   & -9,769         & 0,204 & 0,529 & 5,968 & -0,480 & -0,089 & -1,449 & -3,088 & -2,029          \\
SHORT CALL & NAKED & DELTA BSM & -         & 2\%   & 3,474          & 0,206 & 0,273 & 2,437 & 0,168  & 0,021  & 0,306  & -3,589 & -1,942          \\
SHORT CALL & NAKED & DELTA BSM & -         & 5\%   & 4,854          & 0,108 & 0,171 & 0,623 & 0,450  & 0,128  & 9,944  & -0,635 & -0,045          \\
SHORT CALL &
  NAKED &
  DELTA BSM &
  - &
  10\% &
  2,327 &
  0,069 &
  0,114 &
  \textbf{0,024} &
  0,339 &
  0,069 &
  \textbf{67,508} &
  -0,243 &
  -0,030 \\
SHORT CALL & NAKED & DELTA VG  & -         & 0\%   & -7,634         & 0,185 & 0,439 & 5,968 & -0,413 & -0,072 & -0,918 & -2,770 & -1,968          \\
SHORT CALL & NAKED & DELTA VG  & -         & 2\%   & 2,786          & 0,200 & 0,371 & 2,992 & 0,139  & 0,010  & 0,097  & -3,147 & -1,301          \\
SHORT CALL & NAKED & DELTA VG  & -         & 5\%   & 3,911          & 0,137 & 0,238 & 2,159 & 0,286  & 0,047  & 0,852  & -0,860 & -0,025          \\
SHORT CALL & NAKED & DELTA VG  & -         & 10\%  & 0,840          & 0,022 & 0,022 & 0,036 & 0,375  & 0,142  & 33,424 & -0,087 & -0,003          \\
SHORT CALL & NAKED & VIX       & -         & 0\%   & -3,358         & 0,062 & 0,195 & 5,968 & -0,540 & -0,093 & -0,524 & -1,056 & -0,714          \\
SHORT CALL & NAKED & VIX       & -         & 2\%   & 0,938          & 0,026 & 0,044 & 1,480 & 0,355  & 0,077  & 0,485  & -0,464 & -0,196          \\
SHORT CALL & NAKED & VIX       & -         & 5\%   & 0,501          & 0,006 & 0,008 & 0,179 & 0,779  & 0,477  & 13,375 & -0,064 & -0,006          \\
SHORT CALL & NAKED & VIX       & -         & 10\%  & 0,076          & 0,001 & 0,002 & 0,044 & 0,647  & 0,308  & 5,363  & -0,006 & \textbf{-0,001} \\
SHORT CALL & BSM   & DELTA     & 30        & 0\%   & 0,878          & 0,100 & 0,242 & 2,492 & 0,088  & 0,003  & 0,011  & -1,582 & -0,637          \\
SHORT CALL & BSM   & DELTA     & 30        & 2\%   & 5,428          & 0,090 & 0,193 & 2,611 & 0,602  & 0,169  & 3,519  & -1,332 & -0,462          \\
SHORT CALL & BSM   & DELTA     & 30        & 5\%   & 1,942          & 0,038 & 0,068 & 1,897 & 0,516  & 0,147  & 1,507  & -0,279 & -0,046          \\
SHORT CALL & BSM   & DELTA     & 30        & 10\%  & 0,102          & 0,012 & 0,018 & 3,825 & 0,085  & 0,005  & 0,001  & -0,093 & -0,015          \\
SHORT CALL & BSM   & DELTA     & 130       & 0\%   & 3,347          & 0,113 & 0,231 & 2,492 & 0,297  & 0,043  & 0,578  & -1,724 & -0,693          \\
SHORT CALL & BSM   & DELTA     & 130       & 2\%   & 6,504          & 0,102 & 0,209 & 2,611 & 0,635  & 0,197  & 4,918  & -1,422 & -0,565          \\
SHORT CALL & BSM   & DELTA     & 130       & 5\%   & 3,674          & 0,047 & 0,040 & 1,865 & 0,781  & 0,709  & 13,962 & -0,274 & -0,041          \\
SHORT CALL & BSM   & DELTA     & 130       & 10\%  & 0,404          & 0,012 & 0,018 & 1,516 & 0,329  & 0,075  & 0,200  & -0,083 & -0,016          \\
SHORT CALL & BSM   & DELTA     & SINGLE    & 0\%   & -2,798         & 0,152 & 0,358 & 3,865 & -0,184 & -0,014 & -0,104 & -2,512 & -0,984          \\
SHORT CALL & BSM   & DELTA     & SINGLE    & 2\%   & 3,183          & 0,147 & 0,309 & 2,075 & 0,217  & 0,022  & 0,343  & -2,301 & -0,745          \\
SHORT CALL & BSM   & DELTA     & SINGLE    & 5\%   & 1,799          & 0,059 & 0,126 & 1,901 & 0,305  & 0,043  & 0,411  & -0,355 & -0,038          \\
SHORT CALL & BSM   & DELTA     & SINGLE    & 0\%   & 0,359          & 0,011 & 0,018 & 1,262 & 0,323  & 0,065  & 0,185  & -0,076 & -0,017          \\
SHORT CALL & BSM   & VIX       & 30        & 0\%   & -0,333         & 0,031 & 0,099 & 1,381 & -0,106 & -0,004 & -0,009 & -0,539 & -0,215          \\
SHORT CALL & BSM   & VIX       & 30        & 2\%   & 0,830          & 0,015 & 0,029 & 0,738 & 0,546  & 0,157  & 1,765  & -0,228 & -0,069          \\
SHORT CALL & BSM   & VIX       & 30        & 5\%   & 0,237          & 0,004 & 0,008 & 1,365 & 0,617  & 0,174  & 0,303  & -0,039 & -0,008          \\
SHORT CALL & BSM   & VIX       & 30        & 10\%  & 0,058          & 0,001 & 0,001 & 0,619 & 0,752  & 0,363  & 0,343  & -0,008 & -0,003          \\
SHORT CALL & BSM   & VIX       & 130       & 0\%   & 0,793          & 0,036 & 0,083 & 2,770 & 0,222  & 0,021  & 0,060  & -0,588 & -0,222          \\
SHORT CALL & BSM   & VIX       & 130       & 2\%   & 0,696          & 0,018 & 0,024 & 0,774 & 0,380  & 0,111  & 0,996  & -0,270 & -0,083          \\
SHORT CALL & BSM   & VIX       & 130       & 5\%   & 0,351          & 0,004 & 0,006 & 1,718 & 0,936  & 0,572  & 1,167  & -0,040 & -0,008          \\
SHORT CALL & BSM   & VIX       & 130       & 10\%  & 0,077          & 0,001 & 0,001 & 0,956 & 1,027  & 0,806  & 0,649  & -0,007 & -0,003          \\
SHORT CALL & BSM   & VIX       & SINGLE    & 0\%   & 1,106          & 0,045 & 0,082 & 1,992 & 0,247  & 0,033  & 0,186  & -0,762 & -0,286          \\
SHORT CALL & BSM   & VIX       & SINGLE    & 2\%   & 0,732          & 0,025 & 0,056 & 2,786 & 0,290  & 0,038  & 0,099  & -0,340 & -0,098          \\
SHORT CALL & BSM   & VIX       & SINGLE    & 5\%   & 0,442          & 0,005 & 0,010 & 0,563 & 0,931  & 0,399  & 3,134  & -0,045 & -0,010          \\
SHORT CALL &
  BSM &
  VIX &
  SINGLE &
  10\% &
  0,110 &
  \textbf{0,001} &
  \textbf{0,001} &
  0,528 &
  \textbf{1,582} &
  \textbf{2,453} &
  5,127 &
  \textbf{-0,006} &
  -0,003 \\
SHORT CALL & VG    & DELTA     & 30        & 0\%   & 3,721          & 0,109 & 0,156 & 2,333 & 0,342  & 0,081  & 1,299  & -1,797 & -0,992          \\
SHORT CALL & VG    & DELTA     & 30        & 2\%   & 3,597          & 0,149 & 0,274 & 3,087 & 0,242  & 0,032  & 0,370  & -2,438 & -0,826          \\
SHORT CALL & VG    & DELTA     & 30        & 5\%   & 4,025          & 0,103 & 0,163 & 0,944 & 0,391  & 0,097  & 4,121  & -0,788 & -0,075          \\
SHORT CALL & VG    & DELTA     & 30        & 10\%  & 0,575          & 0,018 & 0,021 & 1,893 & 0,317  & 0,085  & 0,258  & -0,097 & -0,010          \\
SHORT CALL & VG    & DELTA     & 130       & 0\%   & -3,426         & 0,122 & 0,348 & 5,968 & -0,280 & -0,028 & -0,158 & -2,167 & -1,180          \\
SHORT CALL & VG    & DELTA     & 130       & 2\%   & 4,433          & 0,172 & 0,342 & 2,992 & 0,258  & 0,033  & 0,495  & -2,626 & -0,923          \\
SHORT CALL & VG    & DELTA     & 130       & 5\%   & 4,913          & 0,123 & 0,212 & 0,917 & 0,399  & 0,093  & 4,965  & -0,827 & -0,093          \\
SHORT CALL & VG    & DELTA     & 130       & 10\%  & 0,404          & 0,012 & 0,018 & 1,516 & 0,329  & 0,075  & 0,200  & -0,083 & -0,016          \\
SHORT CALL & VG    & DELTA     & SINGLE    & 0\%   & -2,156         & 0,127 & 0,356 & 5,968 & -0,170 & -0,010 & -0,037 & -2,121 & -1,234          \\
SHORT CALL & VG    & DELTA     & SINGLE    & 2\%   & 3,646          & 0,177 & 0,351 & 3,115 & 0,206  & 0,021  & 0,250  & -2,675 & -0,999          \\
SHORT CALL & VG    & DELTA     & SINGLE    & 5\%   & 4,622          & 0,130 & 0,226 & 1,067 & 0,354  & 0,073  & 3,143  & -0,906 & -0,087          \\
SHORT CALL & VG    & DELTA     & SINGLE    & 10\%  & 0,803          & 0,021 & 0,021 & 1,067 & 0,391  & 0,149  & 1,120  & -0,097 & -0,009          \\
SHORT CALL & VG    & VIX       & 30        & 0\%   & 0,445          & 0,028 & 0,050 & 1,440 & 0,157  & 0,014  & 0,044  & -0,481 & -0,247          \\
SHORT CALL & VG    & VIX       & 30        & 2\%   & 0,834          & 0,017 & 0,033 & 0,849 & 0,489  & 0,122  & 1,202  & -0,292 & -0,128          \\
SHORT CALL & VG    & VIX       & 30        & 5\%   & 0,350          & 0,006 & 0,012 & 1,365 & 0,555  & 0,166  & 0,427  & -0,081 & -0,020          \\
SHORT CALL & VG    & VIX       & 30        & 10\%  & 0,051          & 0,001 & 0,002 & 2,194 & 0,520  & 0,110  & 0,026  & -0,012 & -0,003          \\
SHORT CALL & VG    & VIX       & 130       & 0\%   & 0,658          & 0,033 & 0,063 & 2,111 & 0,199  & 0,021  & 0,065  & -0,563 & -0,318          \\
SHORT CALL & VG    & VIX       & 130       & 2\%   & 0,617          & 0,019 & 0,050 & 1,611 & 0,324  & 0,040  & 0,154  & -0,337 & -0,158          \\
SHORT CALL & VG    & VIX       & 130       & 5\%   & 0,473          & 0,007 & 0,010 & 0,817 & 0,686  & 0,324  & 1,878  & -0,095 & -0,025          \\
SHORT CALL & VG    & VIX       & 130       & 10\%  & 0,077          & 0,001 & 0,001 & 0,956 & 1,027  & 0,806  & 0,649  & -0,007 & -0,003          \\
SHORT CALL & VG    & VIX       & SINGLE    & 0\%   & 0,906          & 0,042 & 0,077 & 2,448 & 0,215  & 0,025  & 0,093  & -0,720 & -0,401          \\
SHORT CALL & VG    & VIX       & SINGLE    & 2\%   & 1,924          & 0,021 & 0,049 & 0,595 & 0,911  & 0,358  & 11,585 & -0,366 & -0,187          \\
SHORT CALL & VG    & VIX       & SINGLE    & 5\%   & 0,589          & 0,007 & 0,013 & 0,694 & 0,796  & 0,355  & 3,009  & -0,102 & -0,024          \\
SHORT CALL & VG    & VIX       & SINGLE    & 10\%  & 0,139          & 0,001 & 0,001 & 0,488 & 1,392  & 1,978  & 5,623  & -0,009 & -0,003          \\ \bottomrule
\end{tabular}%
}
\source{The best (i.e. highest or lowest) result in each column is bolded. All strategies with VIX sizing were backtested with risk-factor parameter $\rho = 1.4$. \textit{NAKED} stands for strategies without hedging at all, i.e. only taking short positions in options. \textit{SINGLE} rehedging stands for hedging the portfolio only once a day.}
\end{table}

\subsection{Short Put}
Table \ref{tab:short-put-results} presents the results of the short-put strategies across various combinations. Sixteen parameterizations outperformed the benchmark in terms of IR***, with naked strategies relying on VIX or VG-delta-based sizing demonstrating solid performance.

We begin by comparing naked strategies with different sizing methods. Naked strategies with VIX sizing tend to exhibit lower volatility and drawdowns compared to delta-based sizing methods. For example, the ATM naked strategy using VIX sizing achieves an aRC of 6.568\%, an aSD of 0.071, and an MD of 0.130. This indicates that VIX sizing provides a more stable return profile, supported by a higher IR*** of 20.341, reflecting better risk-adjusted returns. In contrast, naked strategies using delta BSM sizing generally show higher returns but also increased volatility and drawdowns. For instance, the ATM delta BSM strategy yields an aRC of 3.994\% with a significantly higher aSD of 0.279 and an MD of 0.456. This higher risk is reflected in the lower IR*** of 0.166. As the OTM percentage increases, the delta BSM strategy at 5\% OTM improves in terms of return (aRC of 6.403\%) and volatility (aSD of 0.26), but remains more volatile compared to VIX sizing. The delta VG sizing strategies, while offering the potential for higher returns, present mixed performance with varying levels of volatility and drawdowns. At 0\% OTM, the delta VG strategy stands out with an aRC of 13.916\%, but comes with a high aSD of 0.32 and an MD of 0.548. However, as the OTM percentage increases, the performance stabilizes. For instance, at 5\% OTM, the delta VG strategy achieves an aRC of 4.519\% with a lower aSD of 0.101 and an MD of 0.143, suggesting improved risk management.

Several key insights emerge when comparing strategies hedged with the BSM model across different re-hedging frequencies (single, 130 minutes, and 30 minutes) to their naked counterparts. Naked strategies generally exhibit higher returns but also come with significantly higher volatility and drawdowns. For instance, the naked ATM short-put strategy with BSM delta sizing yields an aRC of 3.994\% with an aSD of 0.279. Conversely, implementing a single daily hedging strategy with the BSM model at the same OTM level results in a much lower aRC of 1.950\% but with reduced volatility (aSD of 0.149) and drawdowns (MD of 0.223). This trend of reduced returns coupled with lower risk is consistent across different moneyness levels when comparing naked and hedged strategies. More frequent hedging, such as re-hedging every 130 minutes or every 30 minutes, further stabilizes returns. For example, the delta-based strategy hedged every 130 minutes at 2\% OTM produces an aRC of 5.947\% with an aSD of 0.158 and an MD of 0.216, which is more balanced compared to its naked counterpart. The 30-minute re-hedging strategy at 2\% OTM yields an even higher aRC of 7.890\% with moderate volatility. These frequent hedging strategies effectively mitigate the risk of large drawdowns and high volatility associated with naked strategies. Therefore, while naked strategies may offer higher potential returns, BSM hedged strategies, especially those with more frequent re-hedging, provide a more attractive risk-adjusted performance profile, striking a better balance between returns and risk.

VG-hedged strategies with daily re-hedging (single) do not demonstrate general improvement of risk management compared to naked strategies. For instance, the VG daily hedged strategy with delta sizing at 0\% OTM shows an aRC of 2.978\%, but with a very high aSD of 1.287, and an MD of 0.935. Although the return is lower than the naked counterpart, the strategy significantly reduces volatility and drawdown. Similarly, the VG daily hedged strategy using VIX sizing at 0\% OTM achieves an aRC of 1.580\%, an aSD of 0.041, and an MD of 0.075, offering a much more stable performance profile. Re-hedging every 130 minutes further enhances the risk-adjusted performance of VG-hedged strategies. For instance, the ATM VG 130-minute hedged strategy with delta sizing yields an aRC of 13.127\%, an aSD of 1.536, and an MD of 0.923. This demonstrates a high return similar to the naked strategy but with better risk control. On the other hand, the ATM VG 130-minute hedged strategy with VIX sizing achieves an aRC of 1.677\%, an aSD of 0.038, and an MD of 0.094, indicating enhanced stability and lower risk. VG strategies hedged every 30 minutes provide the most frequent risk adjustments, presumably yielding the most balanced performance among all hedged strategies. Nevertheless, this comes at the cost of decreased profitability, especially for the ATM strategy. VG hedged strategies, particularly those with more frequent re-hedging (such as every 130 minutes or 30 minutes), provide a more balanced performance by mitigating risk. Among the hedged strategies, those using VIX sizing generally offer greater stability and lower risk. Conversely, delta sizing can yield higher returns but also introduces more variability, thus requiring a higher risk tolerance.

Finally, we compare hedged strategies across models and sizing methods. In the case of delta-based sizing, the BSM model provided higher IR*** in 10 out of 12 cases, with the two VG outperformance instances observed for daily hedging. Comparing CVaR and VaR measures also points towards the closed-form model as the better one, indicating its hedging capabilities provide better protection against adverse market movements. On the other hand, results for the VIX sizing approach are not as straightforward as in the delta sizing case. Although the VaR and CVaR measures again point towards the BSM model in the vast majority, the IR*** measure is higher for the VG-hedged strategies in half of the results. Moreover, this particular risk-adjusted metric indicates that the market-calibrated model performed better with strategies closer to at-the-money. 

In summary, naked strategies generally delivered higher returns, particularly those using delta VG sizing. However, these come with increased volatility and drawdowns, which can be undesirable for risk-averse investors. Both BSM and VG models can effectively reduce risk through hedging. More frequent re-hedging (e.g., every 130 or 30 minutes) provides better risk-adjusted performance. The BSM model generally yielded better hedging results, but in risk-adjusted profitability terms, the VG model also provided several desirable strategies.

\begin{table}[!ht]
\caption{Short put strategies results.}
\label{tab:short-put-results}
\resizebox{\columnwidth}{!}{%
\begin{tabular}{@{}cccccccccccccc@{}}
\toprule
\multicolumn{5}{c}{STRATEGY} &
  \multirow{2}{*}{aRC (\%)} &
  \multirow{2}{*}{aSD} &
  \multirow{2}{*}{MD} &
  \multirow{2}{*}{MLD} &
  \multirow{2}{*}{IR} &
  \multirow{2}{*}{IR2} &
  \multirow{2}{*}{IR3} &
  \multirow{2}{*}{CVaR (\%)} &
  \multirow{2}{*}{VaR (\%)} \\ \cmidrule(r){1-5}
OPTIONS   & MODEL & SIZING    & REHEDGING & \%OTM &                 &                &                &       &        &        &        &                 &        \\ \midrule
-         & B\&H  & -         & -         & -     & 9,889           & 0,206          & 0,340          & 1,980 & 0,479  & 0,140  & 6,970  & -1,935          & -3,160 \\ \midrule
SHORT PUT & NAKED & DELTA BSM & -         & 0\%   & 3,994           & 0,279          & 0,456          & 3,016 & 0,143  & 0,013  & 0,166  & -4,893          & -2,802 \\
SHORT PUT & NAKED & DELTA BSM & -         & 2\%   & 4,668           & 0,354          & 0,542          & 2,004 & 0,132  & 0,011  & 0,265  & -6,485          & -2,599 \\
SHORT PUT & NAKED & DELTA BSM & -         & 5\%   & 6,403           & 0,260          & 0,351          & 0,563 & 0,247  & 0,045  & 5,114  & -1,649          & -0,134 \\
SHORT PUT & NAKED & DELTA BSM & -         & 10\%  & 4,138           & 0,305          & 0,378          & 0,040 & 0,136  & 0,015  & 15,451 & -1,073          & -0,017 \\
SHORT PUT & NAKED & DELTA VG  & -         & 0\%   & \textbf{13,916} & 0,320          & 0,548          & 1,722 & 0,435  & 0,110  & 8,916  & -5,560          & -2,440 \\
SHORT PUT & NAKED & DELTA VG  & -         & 2\%   & 4,308           & 0,210          & 0,472          & 3,829 & 0,205  & 0,019  & 0,210  & -3,295          & -1,235 \\
SHORT PUT & NAKED & DELTA VG  & -         & 5\%   & 4,519           & 0,101          & 0,143          & 0,639 & 0,445  & 0,141  & 9,952  & -0,953          & -0,102 \\
SHORT PUT &
  NAKED &
  DELTA VG &
  - &
  10\% &
  2,868 &
  0,079 &
  0,115 &
  \textbf{0,024} &
  0,362 &
  0,090 &
  \textbf{108,648} &
  -0,399 &
  -0,008 \\
SHORT PUT & NAKED & VIX       & -         & 0\%   & 6,568           & 0,071          & 0,130          & 1,508 & 0,926  & 0,467  & 20,341 & -1,279          & -0,722 \\
SHORT PUT & NAKED & VIX       & -         & 2\%   & 3,352           & 0,039          & 0,062          & 1,329 & 0,867  & 0,468  & 11,803 & -0,704          & -0,299 \\
SHORT PUT & NAKED & VIX       & -         & 5\%   & 1,800           & 0,016          & 0,031          & 0,679 & 1,119  & 0,658  & 17,443 & -0,202          & -0,033 \\
SHORT PUT &
  NAKED &
  VIX &
  - &
  10\% &
  0,615 &
  \textbf{0,004} &
  0,009 &
  0,504 &
  \textbf{1,427} &
  \textbf{1,025} &
  12,520 &
  -0,042 &
  \textbf{-0,004} \\
SHORT PUT & BSM   & DELTA     & 30        & 0\%   & 3,603           & 0,113          & 0,172          & 2,171 & 0,320  & 0,067  & 1,113  & -1,438          & -0,602 \\
SHORT PUT & BSM   & DELTA     & 30        & 2\%   & 7,890           & 0,149          & 0,150          & 2,103 & 0,530  & 0,279  & 10,462 & -2,053          & -0,690 \\
SHORT PUT & BSM   & DELTA     & 30        & 5\%   & 7,489           & 0,117          & 0,063          & 2,397 & 0,640  & 0,759  & 23,726 & -0,395          & -0,059 \\
SHORT PUT & BSM   & DELTA     & 30        & 10\%  & 6,965           & 0,154          & 0,059          & 0,774 & 0,454  & 0,536  & 48,244 & -0,202          & -0,012 \\
SHORT PUT & BSM   & DELTA     & 130       & 0\%   & 4,068           & 0,111          & 0,156          & 2,171 & 0,367  & 0,096  & 1,798  & -1,491          & -0,703 \\
SHORT PUT & BSM   & DELTA     & 130       & 2\%   & 5,947           & 0,158          & 0,216          & 2,151 & 0,377  & 0,104  & 2,866  & -2,419          & -0,847 \\
SHORT PUT & BSM   & DELTA     & 130       & 5\%   & 7,485           & 0,098          & 0,061          & 2,381 & 0,761  & 0,936  & 29,419 & -0,400          & -0,068 \\
SHORT PUT & BSM   & DELTA     & 130       & 10\%  & 5,949           & 0,118          & 0,059          & 0,405 & 0,504  & 0,510  & 74,931 & -0,206          & -0,014 \\
SHORT PUT & BSM   & DELTA     & SINGLE    & 0\%   & 1,950           & 0,149          & 0,223          & 1,234 & 0,131  & 0,011  & 0,181  & -2,206          & -0,930 \\
SHORT PUT & BSM   & DELTA     & SINGLE    & 2\%   & -12,968         & 0,245          & 0,644          & 5,881 & -0,529 & -0,107 & -2,351 & -4,455          & -1,541 \\
SHORT PUT & BSM   & DELTA     & SINGLE    & 5\%   & 6,051           & 0,155          & 0,128          & 2,516 & 0,390  & 0,184  & 4,429  & -0,758          & -0,084 \\
SHORT PUT & BSM   & DELTA     & SINGLE    & 10\%  & 8,112           & 0,176          & 0,151          & 1,214 & 0,462  & 0,248  & 16,559 & -0,378          & -0,013 \\
SHORT PUT & BSM   & VIX       & 30        & 0\%   & -1,817          & 0,033          & 0,147          & 2,929 & -0,543 & -0,067 & -0,416 & -0,558          & -0,248 \\
SHORT PUT & BSM   & VIX       & 30        & 2\%   & -0,859          & 0,019          & 0,074          & 5,175 & -0,442 & -0,051 & -0,085 & -0,346          & -0,125 \\
SHORT PUT & BSM   & VIX       & 30        & 5\%   & 0,725           & 0,006          & 0,014          & 1,119 & 1,197  & 0,604  & 3,911  & -0,087          & -0,026 \\
SHORT PUT & BSM   & VIX       & 30        & 10\%  & 0,319           & \textbf{0,004} & 0,008          & 1,444 & 0,755  & 0,306  & 0,676  & \textbf{-0,036} & -0,005 \\
SHORT PUT & BSM   & VIX       & 130       & 0\%   & -1,336          & 0,036          & 0,136          & 2,921 & -0,374 & -0,037 & -0,169 & -0,588          & -0,261 \\
SHORT PUT & BSM   & VIX       & 130       & 2\%   & -0,727          & 0,021          & 0,076          & 5,175 & -0,340 & -0,033 & -0,046 & -0,377          & -0,146 \\
SHORT PUT & BSM   & VIX       & 130       & 5\%   & 0,679           & 0,007          & 0,015          & 1,361 & 1,042  & 0,457  & 2,279  & -0,101          & -0,026 \\
SHORT PUT & BSM   & VIX       & 130       & 10\%  & 0,349           & \textbf{0,004} & 0,009          & 0,976 & 0,834  & 0,333  & 1,192  & -0,039          & -0,006 \\
SHORT PUT & BSM   & VIX       & SINGLE    & 0\%   & -0,776          & 0,045          & 0,165          & 2,083 & -0,171 & -0,008 & -0,030 & -0,755          & -0,289 \\
SHORT PUT & BSM   & VIX       & SINGLE    & 2\%   & -1,309          & 0,029          & 0,087          & 5,766 & -0,446 & -0,067 & -0,152 & -0,516          & -0,163 \\
SHORT PUT & BSM   & VIX       & SINGLE    & 5\%   & 1,007           & 0,009          & 0,018          & 0,754 & 1,176  & 0,672  & 8,970  & -0,119          & -0,027 \\
SHORT PUT & BSM   & VIX       & SINGLE    & 10\%  & 0,407           & 0,005          & \textbf{0,007} & 1,131 & 0,889  & 0,493  & 1,775  & -0,040          & -0,006 \\
SHORT PUT & VG    & DELTA     & 30        & 0\%   & -21,765         & 0,540          & 0,853          & 3,929 & -0,403 & -0,103 & -5,697 & -9,014          & -3,366 \\
SHORT PUT & VG    & DELTA     & 30        & 2\%   & 0,385           & 0,143          & 0,346          & 5,675 & 0,027  & 0,000  & 0,000  & -2,325          & -0,938 \\
SHORT PUT & VG    & DELTA     & 30        & 5\%   & 2,559           & 0,050          & 0,061          & 1,234 & 0,516  & 0,216  & 4,486  & -0,585          & -0,189 \\
SHORT PUT & VG    & DELTA     & 30        & 10\%  & 2,922           & 0,052          & 0,068          & 2,448 & 0,558  & 0,241  & 2,876  & -0,433          & -0,138 \\
SHORT PUT & VG    & DELTA     & 130       & 0\%   & 13,127          & 1,536          & 0,923          & 1,377 & 0,085  & 0,012  & 1,159  & -8,335          & -3,231 \\
SHORT PUT & VG    & DELTA     & 130       & 2\%   & 2,916           & 0,146          & 0,347          & 4,702 & 0,199  & 0,017  & 0,104  & -2,267          & -0,861 \\
SHORT PUT & VG    & DELTA     & 130       & 5\%   & 3,201           & 0,062          & 0,082          & 1,060 & 0,519  & 0,204  & 6,161  & -0,648          & -0,206 \\
SHORT PUT & VG    & DELTA     & 130       & 10\%  & 5,949           & 0,118          & 0,059          & 0,405 & 0,504  & 0,510  & 74,931 & -0,206          & -0,014 \\
SHORT PUT & VG    & DELTA     & SINGLE    & 0\%   & 2,978           & 1,287          & 0,935          & 2,651 & 0,023  & 0,001  & 0,008  & -9,326          & -3,949 \\
SHORT PUT & VG    & DELTA     & SINGLE    & 2\%   & 2,435           & 0,148          & 0,332          & 4,687 & 0,165  & 0,012  & 0,063  & -2,262          & -0,884 \\
SHORT PUT & VG    & DELTA     & SINGLE    & 5\%   & 3,280           & 0,068          & 0,080          & 1,198 & 0,486  & 0,198  & 5,422  & -0,713          & -0,232 \\
SHORT PUT & VG    & DELTA     & SINGLE    & 10\%  & 2,869           & 0,062          & 0,070          & 1,806 & 0,461  & 0,190  & 3,016  & -0,557          & -0,188 \\
SHORT PUT & VG    & VIX       & 30        & 0\%   & 0,436           & 0,034          & 0,079          & 2,083 & 0,129  & 0,007  & 0,015  & -0,611          & -0,333 \\
SHORT PUT & VG    & VIX       & 30        & 2\%   & -0,534          & 0,023          & 0,066          & 2,687 & -0,237 & -0,019 & -0,038 & -0,419          & -0,196 \\
SHORT PUT & VG    & VIX       & 30        & 5\%   & -0,259          & 0,014          & 0,037          & 2,762 & -0,180 & -0,013 & -0,012 & -0,222          & -0,130 \\
SHORT PUT & VG    & VIX       & 30        & 10\%  & -0,946          & 0,012          & 0,066          & 4,980 & -0,808 & -0,115 & -0,218 & -0,186          & -0,115 \\
SHORT PUT & VG    & VIX       & 130       & 0\%   & 1,677           & 0,038          & 0,094          & 2,083 & 0,447  & 0,080  & 0,644  & -0,670          & -0,390 \\
SHORT PUT & VG    & VIX       & 130       & 2\%   & 0,352           & 0,025          & 0,053          & 2,278 & 0,140  & 0,009  & 0,014  & -0,460          & -0,220 \\
SHORT PUT & VG    & VIX       & 130       & 5\%   & 0,118           & 0,016          & 0,030          & 2,762 & 0,074  & 0,003  & 0,001  & -0,248          & -0,138 \\
SHORT PUT & VG    & VIX       & 130       & 10\%  & 0,349           & \textbf{0,004} & 0,009          & 0,976 & 0,834  & 0,333  & 1,192  & -0,039          & -0,006 \\
SHORT PUT & VG    & VIX       & SINGLE    & 0\%   & 1,580           & 0,041          & 0,075          & 2,083 & 0,381  & 0,080  & 0,606  & -0,739          & -0,449 \\
SHORT PUT & VG    & VIX       & SINGLE    & 2\%   & 0,603           & 0,029          & 0,059          & 3,107 & 0,206  & 0,021  & 0,041  & -0,508          & -0,258 \\
SHORT PUT & VG    & VIX       & SINGLE    & 5\%   & 0,373           & 0,017          & 0,028          & 1,159 & 0,219  & 0,029  & 0,093  & -0,265          & -0,155 \\
SHORT PUT & VG    & VIX       & SINGLE    & 10\%  & -0,540          & 0,012          & 0,056          & 4,980 & -0,450 & -0,043 & -0,047 & -0,183          & -0,124 \\ \bottomrule
\end{tabular}%
}
\source{The best (i.e. highest or lowest) result in each column is bolded. All strategies with VIX sizing were backtested with risk-factor parameter $\rho = 1.4$. \textit{NAKED} stands for strategies without hedging at all, i.e. only taking short positions in options. \textit{SINGLE} rehedging stands for hedging the portfolio only once a day.}
\end{table}



\subsection{Short Strangle}
Table \ref{tab:short-strangle-results} presents the performance of short straddles and short strangles across all possible parameterizations. Two naked systems outperformed the benchmark B\&H in terms of aRC. Notably, two naked systems outperformed the B\&H benchmark in terms of aRC. Specifically, both were 5\% OTM strangles employing delta-based sizing — one utilizing the BSM model and the other employing the VG model. Eighteen strategies exhibited superior results compared to the index-tracking benchmark in terms of the IR***.

For the at-the-money (ATM) short straddle strategy with delta sizing based on the BSM model, the annualized return was -1.181\%, with an annualized Standard Deviation (aSD) of 0.123 and an Information Ratio (IR***) of -0.007. When utilizing delta sizing based on the VG model, the aRC decreased further to -2.262\%, IR*** to -0.039, and the aSD increased to 0.134. Conversely, the ATM short straddle strategy employing VIX sizing exhibited different outcomes: an aRC of 3.041\%, an aSD of 0.071, and an IR*** of 1.774.

The performance of the short strangle strategy, involving the sale of out-of-the-money options, varied significantly based on strike distance and sizing method. For instance, the 2\% OTM short strangle with delta sizing using the BSM model yielded an aRC of 4.783\%, an aSD of 0.213, and an IR*** of 0.389. Increasing the strike distance to 5\% OTM improved the aRC to 10.270\%, with an aSD of 0.13 and IR*** of 64.160. The 10\% OTM variant delivered an annualized return of 4.414\%, with an aSD of 0.241. Conversely, when employing delta sizing based on the VG model, the 2\% OTM short strangle achieved an annualized return of 6.636\%, with an aSD of 0.283, an IR*** of 0.672, and a CVaR of -4.931\%. 

The VIX-sized short strangles presented a distinct performance profile. The 2\% OTM variant recorded an aRC of 4.259\%, an aSD of 0.044, an IR*** of 19.402, and a CVaR of -0.791\%. The 5\% OTM variant showed an aRC of 2.318\% and an aSD of 0.017, while the 10\% OTM variant had an annualized return of 0.693\% with an aSD of 0.004.

For the short straddle ATM strategy with BSM delta sizing and a rehedging frequency of 30 minutes, the annualized return was 3.652\%, with an aSD of 0.075, an IR*** of 3.304, and a CVaR of -1.169\%. Decreasing the rehedging frequency to every 130 minutes improved the aRC to 4.876\%, with an aSD of 0.081, and a CVaR of -1.278\%. However, hedging just once daily led to a significant drop in performance, with an aRC of 0.818\%, an aSD of 0.011, and a CVaR of -1.890\%. For the short straddle ATM strategy with VIX sizing, the performance varied with the rehedging frequency. At 30-minute intervals, the aRC was -1.191\%, with an aSD of 4.8\%, and a CVaR of -0.864\%. Rehedging every 130 minutes resulted in an aRC of 0.249\%, an aSD of 0.052, and a CVaR of -0.943\%. With daily rehedging, the aRC was 0.941\%, with an aSD of 0.069, an information ratio of 0.137, and a CVaR of -1.257\%.

In the case of the short strangle strategy with BSM delta sizing at different strike distances, performance was influenced significantly by rehedging frequency. For instance, rehedging every 30 minutes resulted in an aRC of 3.919\% for the 2\% OTM strike, compared to -6.001\% with daily rehedging. VIX sizing also demonstrated varied results across different rehedging frequencies. For instance, in the case of the short strangle strategy with BSM delta sizing and a 2\% OTM strike, rehedging every 30 minutes resulted in an aRC of 3.919\%, an aSD of 0.113, and a CVaR of -2.034\%. When decreasing hedging to 130-minute intervals, the aRC increased to 4.035\%, however, CVaR rose to -2.264\%. Daily rehedging caused a drop in performance, leading to an annualized return of -6.001\%, an aSD of 0.183, and a CVaR of -3.473\%. When using VIX sizing, the aRC with 30-minute rehedging was 4.259\%, an aSD of 0.044, and a CVaR of -0.791\%. For 130-minute intervals, the return was 3.981\%, an aSD of 4.8\%, and a CVaR of -0.862\%. Daily rehedging resulted in an annualized return of 2.947\%, an aSD of 0.051, an information ratio of 0.578, and a CVaR of -0.940\%. Overall, the VIX-based sizing provided more stable results in those cases.

When examining the short strangle strategy with 5\% OTM options to VG-hedged strategy with 30-minute rehedging yielded an aRC of 4.328\%, slightly lower than the naked strategy. However, it demonstrated a lower aSD of 0.255 and a significantly reduced MD of 2.448\%. The VG hedged strategy rehedged every 30 minutes consistently showed improved risk metrics compared to naked strategies. Despite sometimes slightly lower aRC values, the reduction in aSD and CVaR indicates better risk management, which is crucial for maintaining stability during volatile market conditions. The 2\% OTM strangle with delta sizing provided aRC of 7.149\% and IR** of 1.424, which are both better than the naked counterpart. Generally, strategies hedged with the VG every 130 minutes struck a balance between performance and risk, aiming to capture market movements efficiently while avoiding excessive transaction costs associated with more frequent hedging.

While offering simplicity and cost-effectiveness, daily hedging in the VG model generally resulted in lower risk-adjusted returns compared to more frequent hedging intervals. However, when comparing the delta-sizing strategies with options 2\%, 5\%, and 10\% out-of-the-money, the differences in the risk-adjusted metrics are relatively small, hence it is worth considering adjusting the position only once a day. 

To make the analysis complete, we also need to look at the comparison of hedged strategies between the two models. When comparing the delta sizing, the BSM model performs much better and delivers higher IR*** in 10 out of 12 backtest combinations. Moreover, it provides better VaR and CVaR measures in all considered delta-based strategies, making it a better choice for those applications. On the other hand, VIX-sizing results showed that the VG model is capable of providing higher IR*** in half of the backtests. Nevertheless, the closed-form model again proved itself useful by showcasing better hedging capabilities and providing better portfolio VaR and CVaR across all backtests.

The comparison of naked and BSM-hedged strategies across different hedging time frames and sizing methods highlights significant differences in performance. Naked strategies tend to offer higher returns but come with greater risk. The BSM-hedged strategies, particularly those with frequent rehedging, provide a more stable return profile with reduced risk, though the frequency of rehedging and sizing method play crucial roles in determining their efficacy. The VG-hedged strategies are capable of providing profits but come with relatively higher volatility. In conclusion, while naked strategies offer the potential for higher returns, they come at the expense of increased volatility and drawdowns. Implementing the VG model for hedging, particularly with more frequent rehedging intervals and delta-based sizing, emerges as a prudent approach for balancing risk and return in quantitative option writing strategies. The best hedging results are however achieved using the Black-Scholes-Merton model. 


\begin{table}[!ht]
\caption{Short strangle strategies results.}
\label{tab:short-strangle-results}
\resizebox{\columnwidth}{!}{%
\begin{tabular}{@{}cccccccccccccc@{}}
\toprule
\multicolumn{5}{c}{STRATEGY} &
  \multirow{2}{*}{aRC (\%)} &
  \multirow{2}{*}{aSD} &
  \multirow{2}{*}{MD} &
  \multirow{2}{*}{MLD} &
  \multirow{2}{*}{IR} &
  \multirow{2}{*}{IR2} &
  \multirow{2}{*}{IR3} &
  \multirow{2}{*}{CVaR (\%)} &
  \multirow{2}{*}{VaR (\%)} \\ \cmidrule(r){1-5}
OPTIONS        & MODEL & SIZING    & REHEDGING & \%OTM &        &                &       &       &                &                &        &                 &        \\ \midrule
-              & B\&H  & -         & -         & -     & 9,889  & 0,206          & 0,340 & 1,980 & 0,479          & 0,140          & 6,970  & -1,935          & -3,160 \\ \midrule
SHORT STRADDLE & NAKED & DELTA BSM & -         & 0\%   & -1,181 & 0,123          & 0,319 & 5,952 & -0,096         & -0,004         & -0,007 & -2,060          & -1,229 \\
SHORT STRANGLE & NAKED & DELTA BSM & -         & 2\%   & 4,783  & 0,213          & 0,414 & 3,198 & 0,225          & 0,026          & 0,389  & -4,000          & -2,017 \\
SHORT STRANGLE & NAKED & DELTA BSM & -         & 5\%   & 10,270 & 0,130          & 0,189 & 0,687 & 0,789          & 0,429          & 64,160 & -1,085          & -0,145 \\
SHORT STRANGLE & NAKED & DELTA BSM & -         & 10\%  & 4,414  & 0,241          & 0,349 & 0,083 & 0,183          & 0,023          & 12,295 & -1,015          & -0,017 \\
SHORT STRADDLE & NAKED & DELTA VG  & -         & 0\%   & -2,262 & 0,134          & 0,371 & 5,952 & -0,169         & -0,010         & -0,039 & -2,313          & -1,327 \\
SHORT STRANGLE & NAKED & DELTA VG  & -         & 2\%   & 6,636  & 0,283          & 0,518 & 2,964 & 0,234          & 0,030          & 0,672  & -4,931          & -2,242 \\
SHORT STRANGLE &
  NAKED &
  DELTA VG &
  - &
  5\% &
  \textbf{10,409} &
  0,213 &
  0,265 &
  1,214 &
  0,488 &
  0,192 &
  16,453 &
  -1,924 &
  -0,256 \\
SHORT STRANGLE &
  NAKED &
  DELTA VG &
  - &
  10\% &
  6,303 &
  0,244 &
  0,336 &
  \textbf{0,032} &
  0,259 &
  0,049 &
  \textbf{96,347} &
  -0,935 &
  -0,014 \\
SHORT STRADDLE & NAKED & VIX       & -         & 0\%   & 3,041  & 0,071          & 0,100 & 2,246 & 0,431          & 0,131          & 1,774  & -1,237          & -0,717 \\
SHORT STRANGLE & NAKED & VIX       & -         & 2\%   & 4,259  & 0,044          & 0,079 & 1,147 & 0,971          & 0,522          & 19,402 & -0,791          & -0,392 \\
SHORT STRANGLE & NAKED & VIX       & -         & 5\%   & 2,318  & 0,017          & 0,030 & 0,488 & 1,370          & 1,053          & 50,026 & -0,222          & -0,052 \\
SHORT STRANGLE & NAKED & VIX       & -         & 10\%  & 0,693  & 0,004          & 0,009 & 0,405 & \textbf{1,565} & 1,271          & 21,759 & -0,044          & -0,005 \\
SHORT STRADDLE & BSM   & DELTA     & 30        & 0\%   & 3,652  & 0,075          & 0,144 & 1,373 & 0,489          & 0,124          & 3,304  & -1,169          & -0,573 \\
SHORT STRANGLE & BSM   & DELTA     & 30        & 2\%   & 3,919  & 0,113          & 0,159 & 2,520 & 0,348          & 0,085          & 1,329  & -2,034          & -0,943 \\
SHORT STRANGLE & BSM   & DELTA     & 30        & 5\%   & 7,300  & 0,074          & 0,065 & 1,230 & 0,986          & 1,108          & 65,741 & -0,577          & -0,124 \\
SHORT STRANGLE & BSM   & DELTA     & 30        & 10\%  & 6,651  & 0,117          & 0,122 & 0,361 & 0,568          & 0,311          & 57,272 & -0,361          & -0,015 \\
SHORT STRADDLE & BSM   & DELTA     & 130       & 0\%   & 4,876  & 0,081          & 0,090 & 1,254 & 0,604          & 0,328          & 12,756 & -1,278          & -0,622 \\
SHORT STRANGLE & BSM   & DELTA     & 130       & 2\%   & 4,035  & 0,128          & 0,190 & 2,258 & 0,316          & 0,067          & 1,200  & -2,264          & -1,025 \\
SHORT STRANGLE & BSM   & DELTA     & 130       & 5\%   & 7,963  & 0,067          & 0,066 & 1,897 & 1,181          & \textbf{1,421} & 59,669 & -0,568          & -0,112 \\
SHORT STRANGLE & BSM   & DELTA     & 130       & 10\%  & 6,689  & 0,101          & 0,127 & 0,361 & 0,660          & 0,347          & 64,206 & -0,397          & -0,016 \\
SHORT STRADDLE & BSM   & DELTA     & SINGLE    & 0\%   & 0,818  & 0,110          & 0,181 & 2,083 & 0,075          & 0,003          & 0,013  & -1,890          & -0,888 \\
SHORT STRANGLE & BSM   & DELTA     & SINGLE    & 2\%   & -6,001 & 0,183          & 0,433 & 5,897 & -0,327         & -0,045         & -0,462 & -3,473          & -1,477 \\
SHORT STRANGLE & BSM   & DELTA     & SINGLE    & 5\%   & 7,705  & 0,097          & 0,116 & 1,079 & 0,795          & 0,528          & 37,659 & -0,875          & -0,153 \\
SHORT STRANGLE & BSM   & DELTA     & SINGLE    & 10\%  & 5,703  & 0,153          & 0,259 & 1,214 & 0,373          & 0,082          & 3,850  & -0,625          & -0,019 \\
SHORT STRADDLE & BSM   & VIX       & 30        & 0\%   & -1,191 & 0,048          & 0,194 & 2,369 & -0,250         & -0,015         & -0,077 & -0,864          & -0,399 \\
SHORT STRANGLE & BSM   & VIX       & 30        & 2\%   & 0,305  & 0,027          & 0,079 & 2,095 & 0,115          & 0,004          & 0,006  & -0,470          & -0,209 \\
SHORT STRANGLE & BSM   & VIX       & 30        & 5\%   & 1,010  & 0,007          & 0,015 & 0,810 & 1,370          & 0,952          & 11,879 & -0,111          & -0,030 \\
SHORT STRANGLE & BSM   & VIX       & 30        & 10\%  & 0,399  & 0,004          & 0,007 & 1,087 & 0,978          & 0,543          & 1,992  & \textbf{-0,037} & -0,005 \\
SHORT STRADDLE & BSM   & VIX       & 130       & 0\%   & 0,249  & 0,052          & 0,149 & 2,921 & 0,048          & 0,001          & 0,001  & -0,943          & -0,453 \\
SHORT STRANGLE & BSM   & VIX       & 130       & 2\%   & 0,208  & 0,030          & 0,072 & 2,095 & 0,069          & 0,002          & 0,002  & -0,526          & -0,222 \\
SHORT STRANGLE & BSM   & VIX       & 130       & 5\%   & 1,075  & 0,008          & 0,016 & 0,774 & 1,392          & 0,933          & 12,963 & -0,126          & -0,035 \\
SHORT STRANGLE & BSM   & VIX       & 130       & 10\%  & 0,443  & \textbf{0,004} & 0,008 & 0,766 & 1,098          & 0,635          & 3,672  & -0,039          & -0,005 \\
SHORT STRADDLE & BSM   & VIX       & SINGLE    & 0\%   & 0,941  & 0,069          & 0,164 & 2,083 & 0,137          & 0,008          & 0,036  & -1,257          & -0,548 \\
SHORT STRANGLE & BSM   & VIX       & SINGLE    & 2\%   & -0,132 & 0,043          & 0,077 & 2,976 & -0,031         & -0,001         & 0,000  & -0,705          & -0,286 \\
SHORT STRANGLE & BSM   & VIX       & SINGLE    & 5\%   & 1,471  & 0,010          & 0,017 & 0,512 & 1,449          & 1,223          & 35,163 & -0,146          & -0,042 \\
SHORT STRANGLE &
  BSM &
  VIX &
  SINGLE &
  10\% &
  0,521 &
  0,005 &
  \textbf{0,007} &
  0,690 &
  1,147 &
  0,852 &
  6,424 &
  -0,040 &
  \textbf{-0,005} \\
SHORT STRADDLE & VG    & DELTA     & 30        & 0\%   & -0,613 & 0,114          & 0,249 & 5,960 & -0,054         & -0,001         & -0,001 & -2,021          & -0,892 \\
SHORT STRANGLE & VG    & DELTA     & 30        & 2\%   & 4,328  & 0,255          & 0,433 & 2,448 & 0,169          & 0,017          & 0,299  & -4,459          & -2,240 \\
SHORT STRANGLE & VG    & DELTA     & 30        & 5\%   & 4,751  & 0,165          & 0,209 & 1,651 & 0,288          & 0,065          & 1,882  & -1,915          & -0,517 \\
SHORT STRANGLE & VG    & DELTA     & 30        & 10\%  & 5,645  & 0,167          & 0,230 & 3,730 & 0,338          & 0,083          & 1,256  & -0,873          & -0,254 \\
SHORT STRADDLE & VG    & DELTA     & 130       & 0\%   & -5,747 & 0,151          & 0,417 & 5,952 & -0,381         & -0,052         & -0,507 & -2,731          & -1,303 \\
SHORT STRANGLE & VG    & DELTA     & 130       & 2\%   & 7,149  & 0,250          & 0,436 & 2,353 & 0,286          & 0,047          & 1,424  & -4,404          & -2,242 \\
SHORT STRANGLE & VG    & DELTA     & 130       & 5\%   & 5,417  & 0,185          & 0,228 & 1,651 & 0,293          & 0,070          & 2,285  & -2,036          & -0,512 \\
SHORT STRANGLE & VG    & DELTA     & 130       & 10\%  & 6,689  & 0,101          & 0,127 & 0,361 & 0,660          & 0,347          & 64,206 & -0,397          & -0,016 \\
SHORT STRADDLE & VG    & DELTA     & SINGLE    & 0\%   & -3,061 & 0,178          & 0,454 & 5,952 & -0,172         & -0,012         & -0,060 & -3,167          & -1,529 \\
SHORT STRANGLE & VG    & DELTA     & SINGLE    & 2\%   & 5,415  & 0,254          & 0,459 & 2,492 & 0,213          & 0,025          & 0,547  & -4,503          & -2,232 \\
SHORT STRANGLE & VG    & DELTA     & SINGLE    & 5\%   & 9,065  & 0,196          & 0,230 & 1,214 & 0,463          & 0,182          & 13,591 & -1,903          & -0,419 \\
SHORT STRANGLE & VG    & DELTA     & SINGLE    & 10\%  & 5,449  & 0,194          & 0,268 & 2,488 & 0,280          & 0,057          & 1,248  & -0,954          & -0,216 \\
SHORT STRADDLE & VG    & VIX       & 30        & 0\%   & 1,150  & 0,063          & 0,170 & 2,365 & 0,181          & 0,012          & 0,060  & -1,126          & -0,521 \\
SHORT STRANGLE & VG    & VIX       & 30        & 2\%   & 0,373  & 0,033          & 0,085 & 2,266 & 0,114          & 0,005          & 0,008  & -0,587          & -0,311 \\
SHORT STRANGLE & VG    & VIX       & 30        & 5\%   & 0,096  & 0,014          & 0,030 & 2,762 & 0,067          & 0,002          & 0,001  & -0,224          & -0,127 \\
SHORT STRANGLE & VG    & VIX       & 30        & 10\%  & -0,868 & 0,012          & 0,065 & 4,980 & -0,747         & -0,099         & -0,173 & -0,183          & -0,113 \\
SHORT STRADDLE & VG    & VIX       & 130       & 0\%   & 3,638  & 0,077          & 0,149 & 2,290 & 0,473          & 0,115          & 1,830  & -1,349          & -0,725 \\
SHORT STRANGLE & VG    & VIX       & 130       & 2\%   & 1,345  & 0,038          & 0,101 & 2,163 & 0,358          & 0,048          & 0,296  & -0,688          & -0,357 \\
SHORT STRANGLE & VG    & VIX       & 130       & 5\%   & 0,552  & 0,016          & 0,027 & 1,508 & 0,347          & 0,072          & 0,264  & -0,244          & -0,131 \\
SHORT STRANGLE & VG    & VIX       & 130       & 10\%  & 0,443  & \textbf{0,004} & 0,008 & 0,766 & 1,098          & 0,635          & 3,672  & -0,039          & -0,005 \\
SHORT STRADDLE & VG    & VIX       & SINGLE    & 0\%   & 4,866  & 0,088          & 0,174 & 1,873 & 0,550          & 0,154          & 4,002  & -1,571          & -0,780 \\
SHORT STRANGLE & VG    & VIX       & SINGLE    & 2\%   & 2,367  & 0,043          & 0,091 & 1,762 & 0,550          & 0,143          & 1,920  & -0,762          & -0,399 \\
SHORT STRANGLE & VG    & VIX       & SINGLE    & 5\%   & 1,010  & 0,017          & 0,021 & 0,790 & 0,609          & 0,289          & 3,700  & -0,252          & -0,127 \\
SHORT STRANGLE & VG    & VIX       & SINGLE    & 10\%  & -0,396 & 0,012          & 0,051 & 4,571 & -0,336         & -0,026         & -0,023 & -0,178          & -0,121 \\ \bottomrule
\end{tabular}%
}
\source{The best (i.e. highest or lowest) result in each column is bolded. All strategies with VIX sizing were backtested with risk-factor parameter $\rho = 1.4$. \textit{NAKED} stands for strategies without hedging at all, i.e. only taking short positions in options. \textit{SINGLE} rehedging stands for hedging the portfolio only once a day.}
\end{table}

\section{Conclusions}
This research provided an overview of model-based volatility trading with S\&P500 index options, exploring option-writing strategies with varying moneyness levels. Two sizing methods, based on delta and the VIX Index, were designed. All hedged strategies used portfolio delta approximations from the BSM model with implied volatility or from the Variance-Gamma model calibrated to the market. Buy-and-hold and naked strategies were presented as benchmarks. The analysis leveraged 1-minute data of the S\&P500 Index and corresponding options from the CBOE, alongside metrics like VIX index for volatility and risk-free rates for pricing models.

The findings reveal distinct performance patterns across different option writing strategies. Short call strategies, particularly those employing VIX-based sizing, demonstrated stability with minimal drawdowns compared to delta-based strategies. Strategies hedged with the Black-Scholes-Merton (BSM) model, especially those with more frequent rehedging intervals, consistently delivered improved risk-adjusted returns and lower volatility. In contrast, the VG-hedged strategies exhibited greater exposure to risk.

Similarly, in short-put strategies, VIX-sized strategies generally exhibited lower volatility and drawdowns than their delta-sized counterparts. BSM-hedged strategies, again with more frequent rehedging, provided a balanced approach with reduced risk, highlighting their effectiveness in managing downside risk while maintaining competitive returns.

The analysis of short strangle strategies underscored the impact of strike distance, sizing method, and rehedging frequency on performance. Delta-based strategies with BSM model hedging showed varied performance across different strike distances and rehedging intervals, demonstrating the potential for stable returns with controlled risk. 

Comparative analysis between the BSM and VG models revealed that the BSM model consistently outperformed in terms of risk measures such as Conditional Value at Risk (CVaR) and Value at Risk (VaR) across most strategies. The VG model showed competitive performance in a few systems, mainly the VIX-sized strategies, highlighting its suitability under specific circumstances.

In conclusion, while naked strategies offered higher potential returns, they also posed greater risks in terms of volatility and drawdowns. Hedged strategies, particularly those using the BSM model with optimal rehedging intervals, emerged as effective in balancing risk and return. Investors and practitioners may consider implementing these strategies to achieve stable returns while managing market risk effectively.

Looking forward, future research could delve deeper into optimizing rehedging intervals and exploring alternative hedging models to further refine risk-adjusted returns in option trading strategies. 

\clearpage
\bibliography{main}
\clearpage
\appendix

\section{Equity Plots of Naked Strategies}

\subsection{Short Calls}


\begin{figure}[!h]
\centering
\subfloat[\centering Naked Short Calls with BSM Delta Sizing]{{\includegraphics[width=0.55\columnwidth]{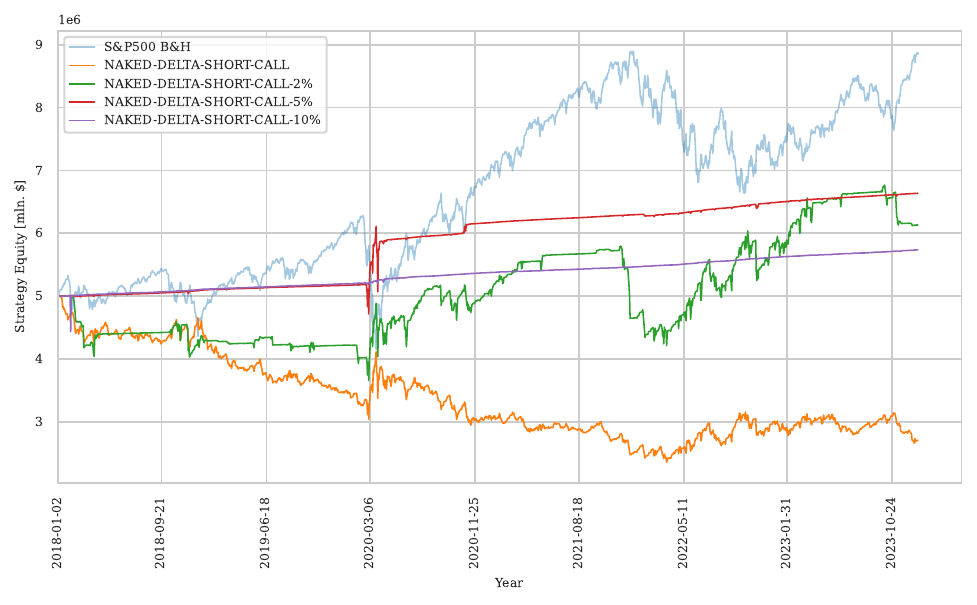} }}\\
\subfloat[\centering Naked Short Calls with VG Delta Sizing]{{\includegraphics[width=0.55\columnwidth]{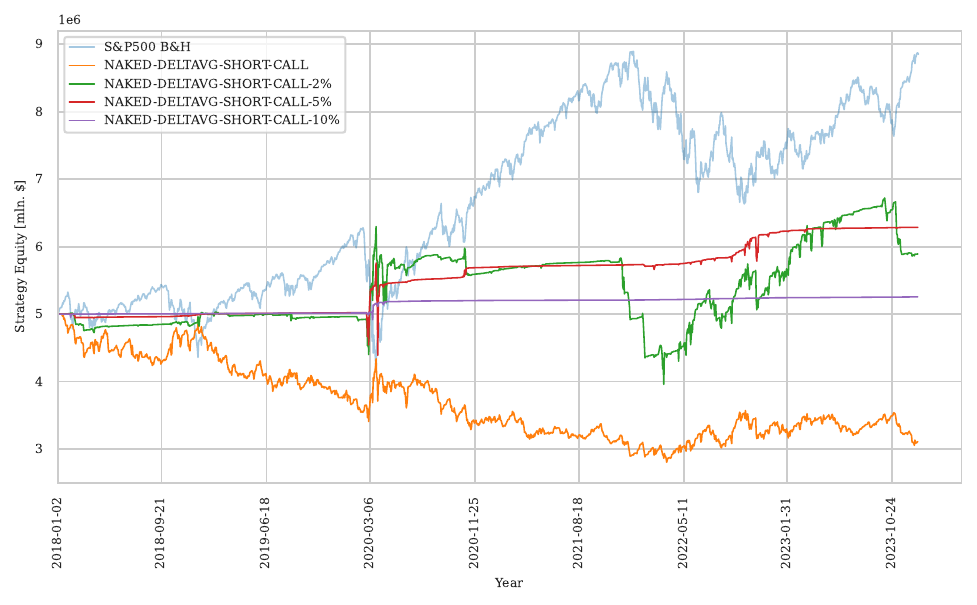} }}\\
\subfloat[\centering Naked Short Calls with VIX Sizing]{{\includegraphics[width=0.55\columnwidth]{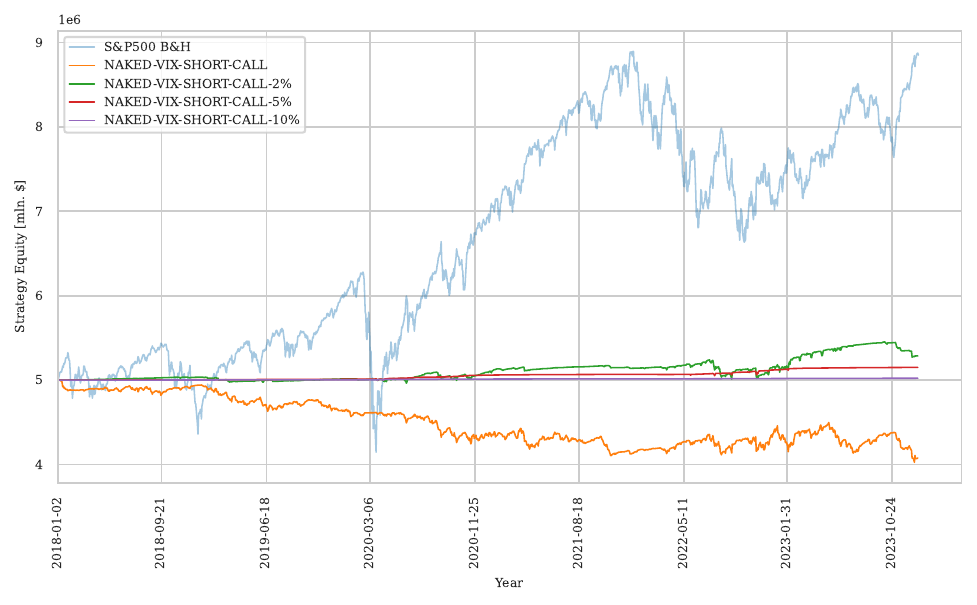} }}
\caption{Naked Short Call Strategies Equity Lines Comparison}%
\label{fig:naked-short-calls}%
\end{figure}

\clearpage
\subsection{Short Puts}
\begin{figure}[!h]
\centering
\subfloat[\centering Naked Short Puts with BSM Delta Sizing]{{\includegraphics[width=0.6\columnwidth]{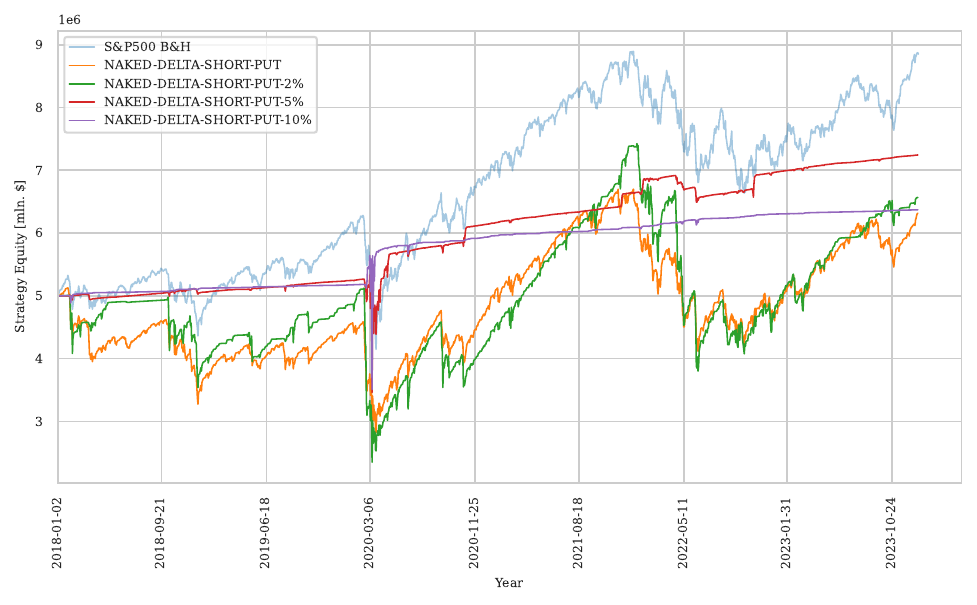} }}\\
\subfloat[\centering Naked Short Puts with VG Delta Sizing]{{\includegraphics[width=0.6\columnwidth]{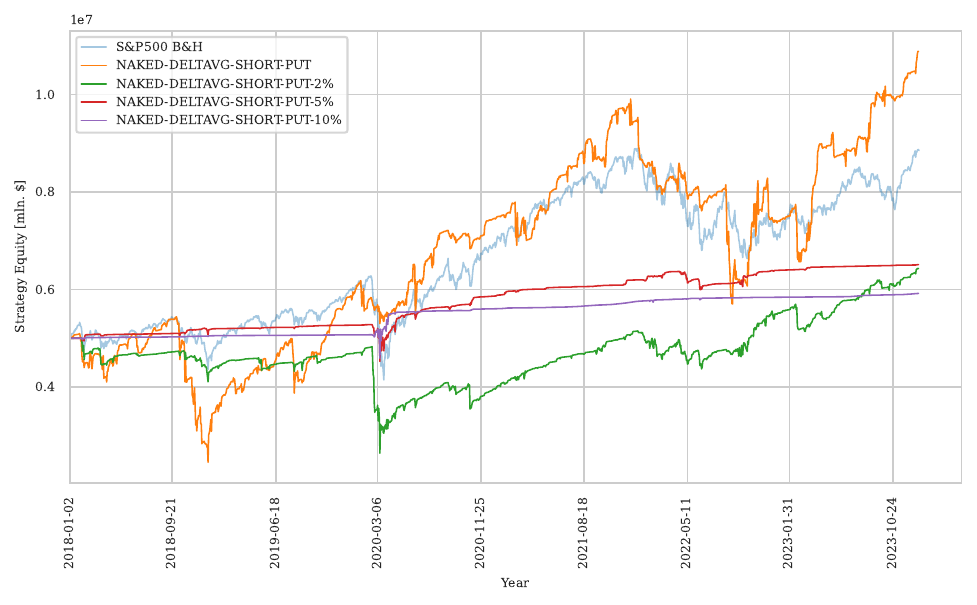} }}\\
\subfloat[\centering Naked Short Puts with VIX Sizing]{{\includegraphics[width=0.6\columnwidth]{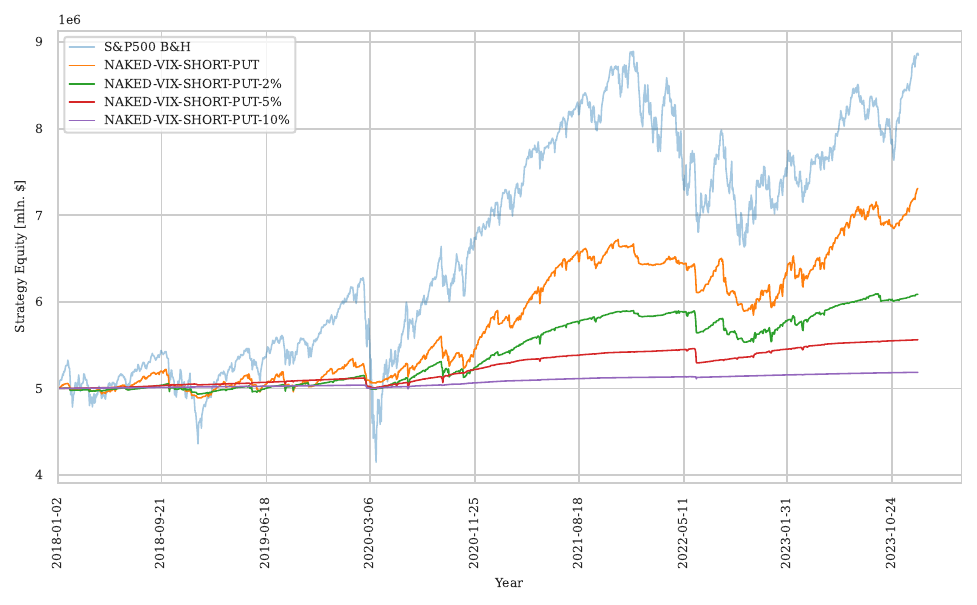} }}
\caption{Naked Short Put Strategies Equity Lines Comparison}%
\label{fig:naked-short-puts}%
\end{figure}

\clearpage
\subsection{Short Strangles}
\begin{figure}[!h]
\centering
\subfloat[\centering Naked Short Strangles with BSM Delta Sizing]{{\includegraphics[width=0.6\columnwidth]{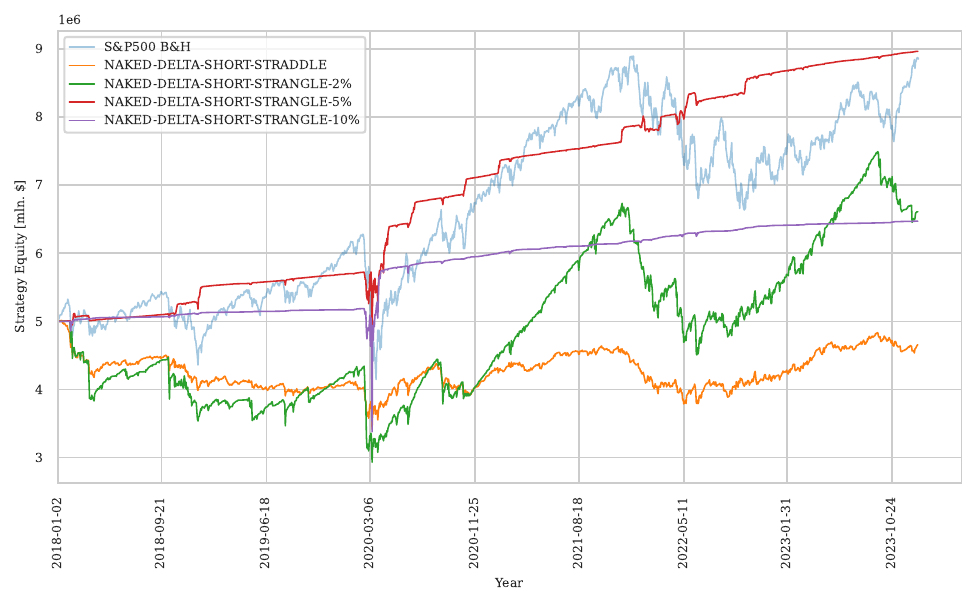} }} \\
\subfloat[\centering Naked Short Strangles with VG Delta Sizing]{{\includegraphics[width=0.6\columnwidth]{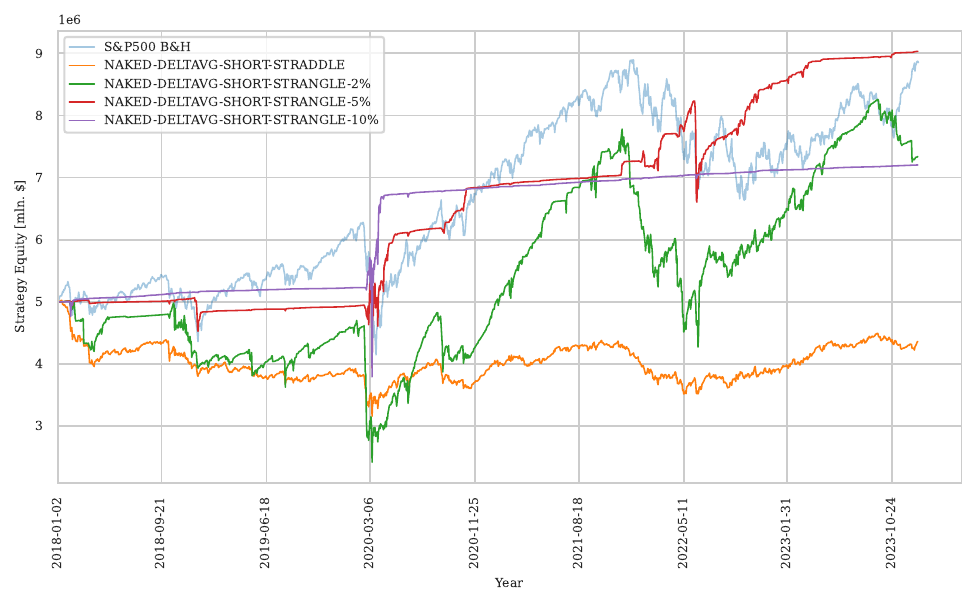} }}\\
\subfloat[\centering Naked Short Strangles with VIX Sizing]{{\includegraphics[width=0.6\columnwidth]{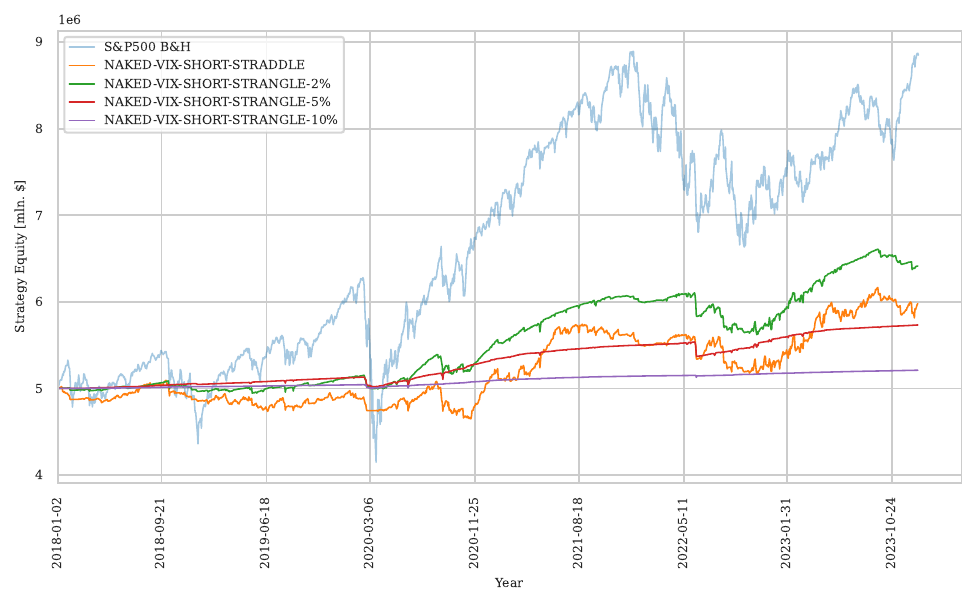}}}
\caption{Naked Short Strangle Strategies Equity Lines Comparison}%
\label{fig:naked-short-strangles}%
\end{figure}

\section{Equity Plots of Strategies with Daily Hedging}
\subsection{Short Calls}
\begin{figure}[!h]
\centering
\subfloat[\centering Daily Hedged Short Calls with Delta Sizing]{{\includegraphics[width=0.88\columnwidth]{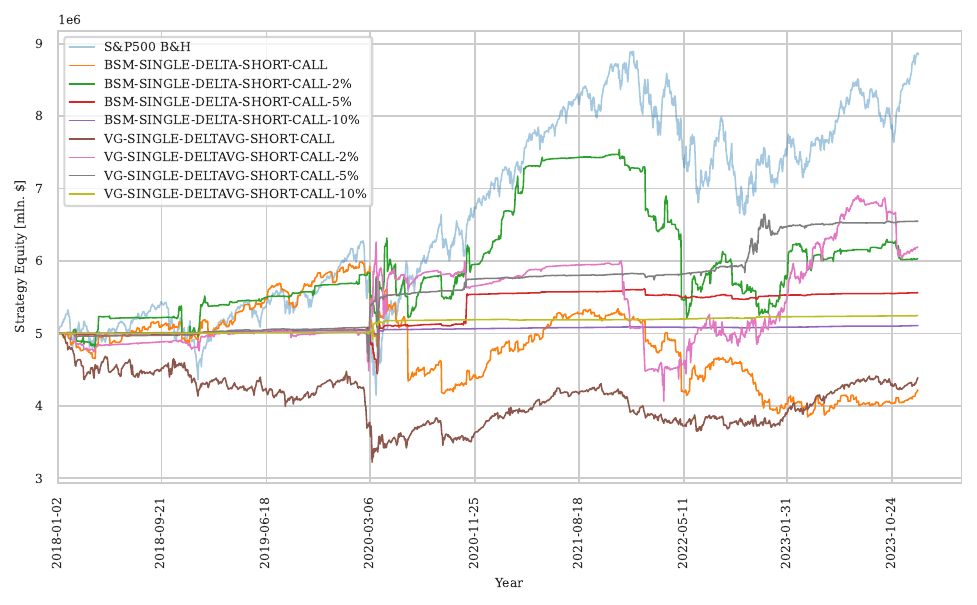} }}\\
\subfloat[\centering Daily Hedged Short Calls with VIX Sizing]{{\includegraphics[width=0.88\columnwidth]{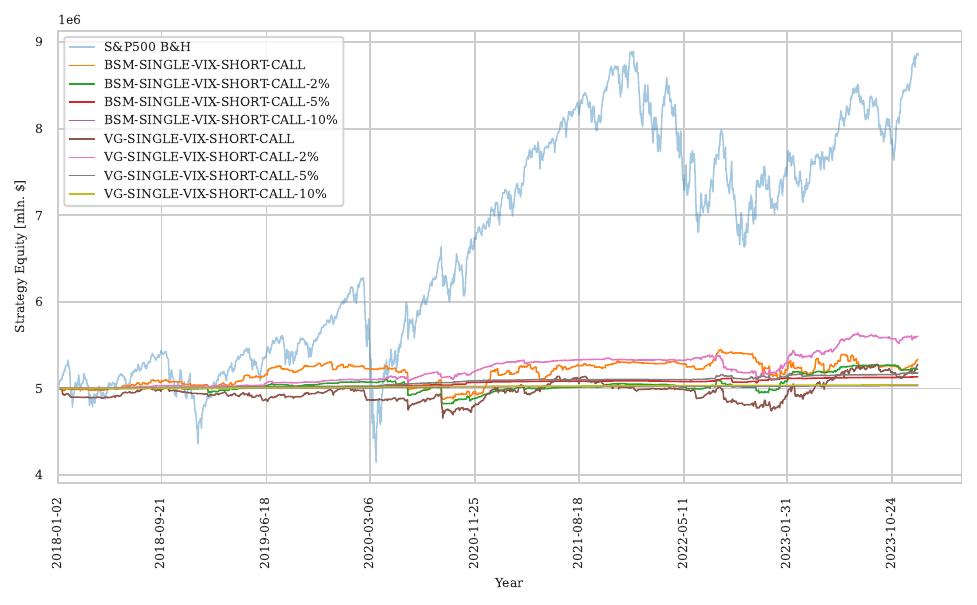} }}
\caption{Daily Hedged Short Call Strategies Equity Lines Comparison}%
\label{fig:single-short-calls}%
\end{figure}

\clearpage
\subsection{Short Puts}
\begin{figure}[!h]
\centering
\subfloat[\centering Daily Hedged Short Puts with Delta Sizing]{{\includegraphics[width=0.88\columnwidth]{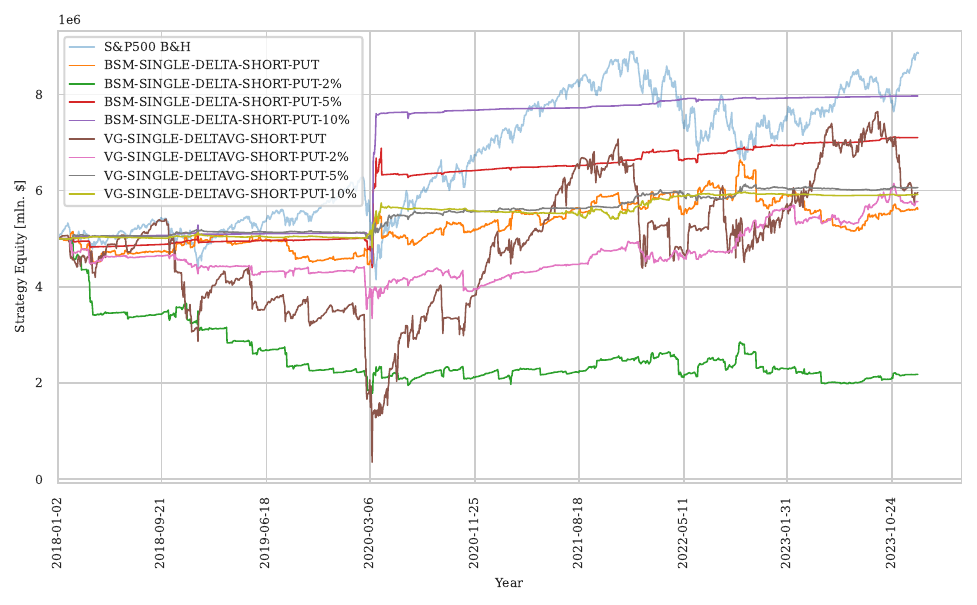} }}\\
\subfloat[\centering Daily Hedged Short Puts with VIX Sizing]{{\includegraphics[width=0.88\columnwidth]{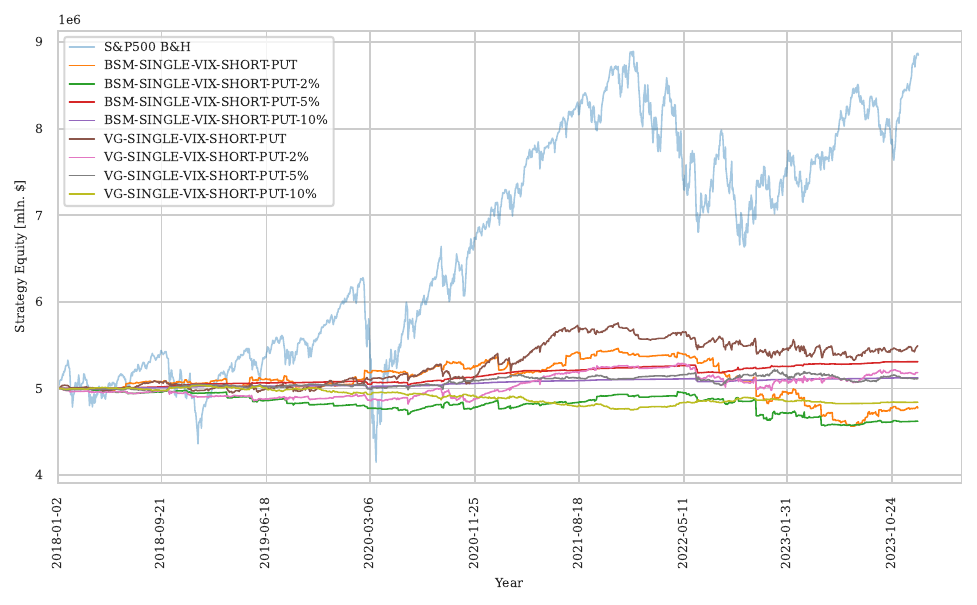} }}
\caption{Daily Hedged Short Put Strategies Equity Lines Comparison}%
\label{fig:single-short-puts}%
\end{figure}

\clearpage
\subsection{Short Strangles}
\begin{figure}[!h]
\centering
\subfloat[\centering Daily Hedged Short Strangles with Delta Sizing]{{\includegraphics[width=0.88\columnwidth]{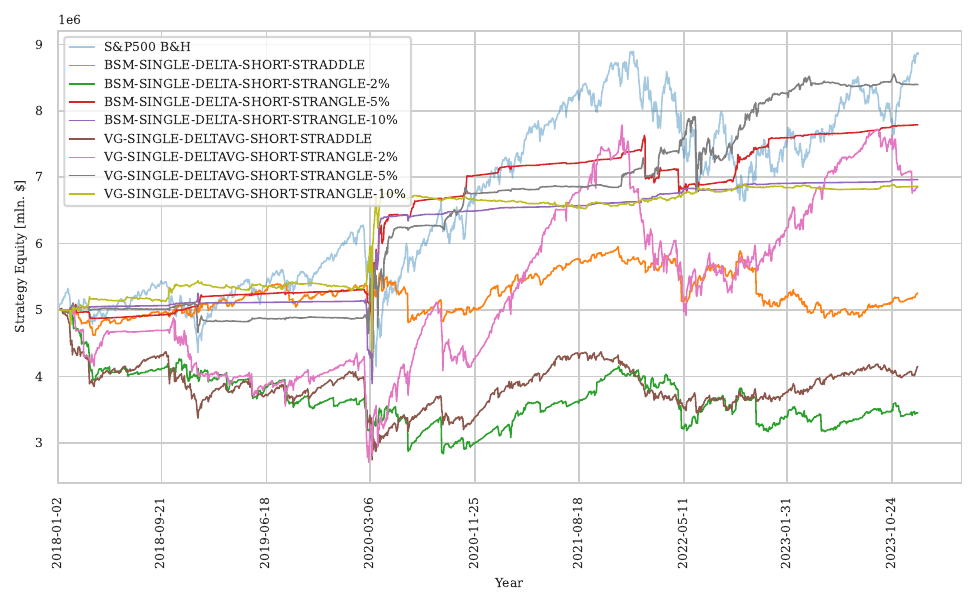} }}\\
\subfloat[\centering Daily Hedged Short Strangles with VIX Sizing]{{\includegraphics[width=0.88\columnwidth]{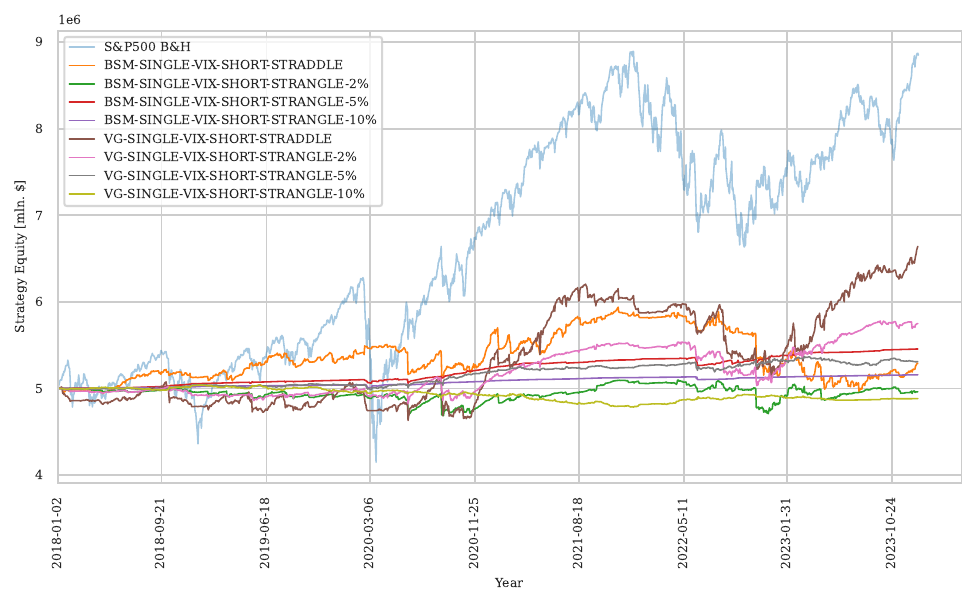} }}
\caption{Daily Hedged Short Strangle Strategies Equity Lines Comparison}%
\label{fig:single-short-strangles}%
\end{figure}

\section{Equity Plots of Strategies with Intraday Hedging}
\subsection{Short Calls}
\subsubsection{Hedging Every 130 Minutes}
\begin{figure}[!h]
\centering
\subfloat[\centering Short Calls with Delta Sizing Hedged Every 130 Minutes]{{\includegraphics[width=0.8\columnwidth]{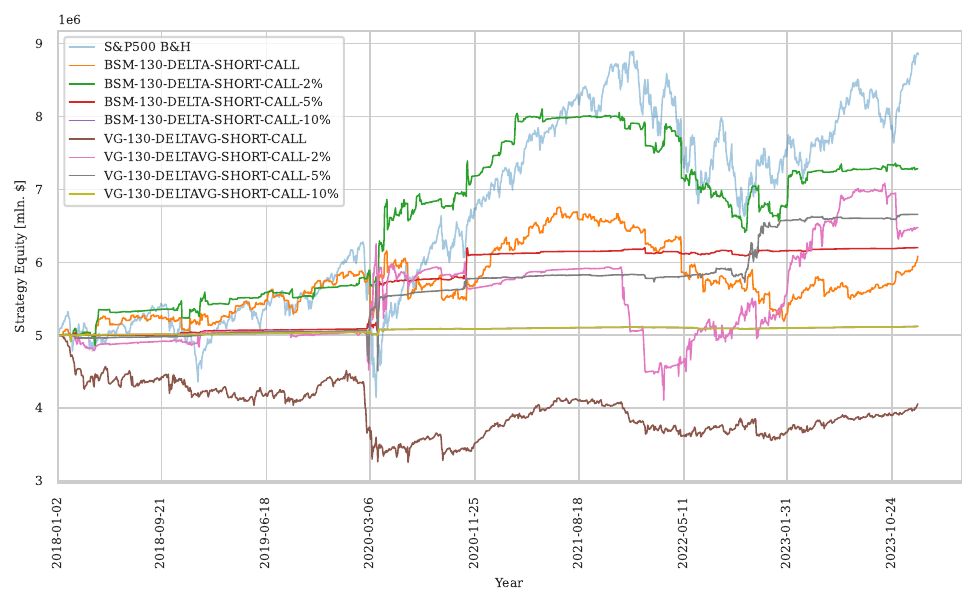} }}\\
\subfloat[\centering Short Calls with VIX Sizing Hedged Every 130 Minutes]{{\includegraphics[width=0.8\columnwidth]{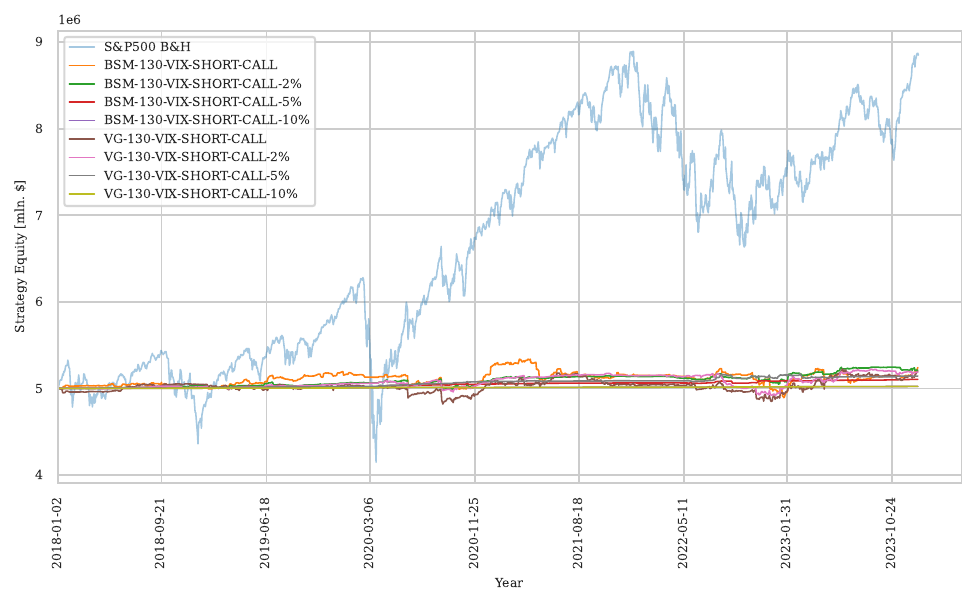} }}
\caption{Intraday Hedged (130 Minutes) Short Call Strategies Equity Lines Comparison}%
\label{fig:130-short-calls}%
\end{figure}

\clearpage
\subsubsection{Hedging Every 30 Minutes}
\begin{figure}[!h]
\centering
\subfloat[\centering Short Calls with Delta Sizing Hedged Every 30 Minutes]{{\includegraphics[width=0.88\columnwidth]{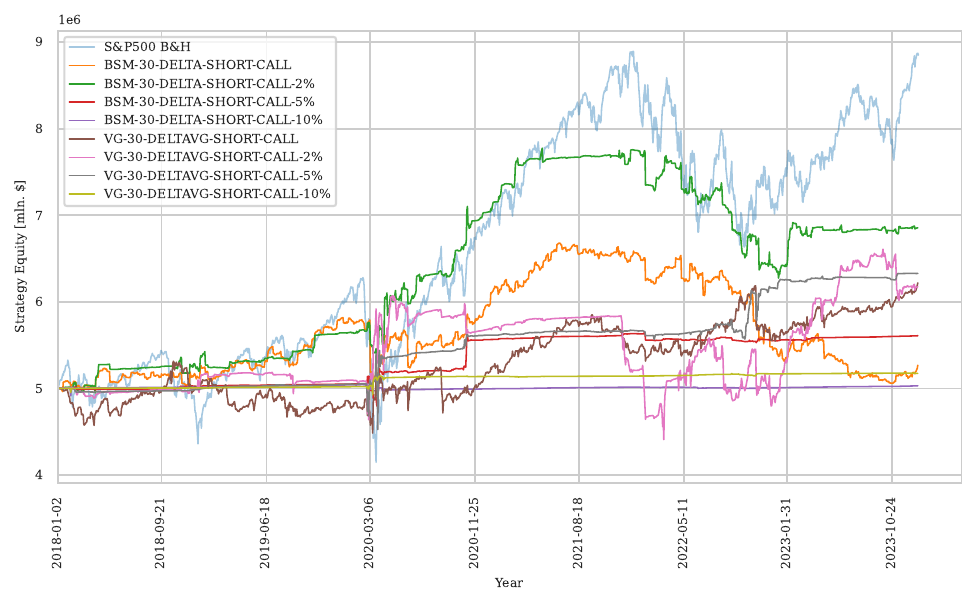} }}\\
\subfloat[\centering Short Calls with VIX Sizing Hedged Every 30 Minutes]{{\includegraphics[width=0.88\columnwidth]{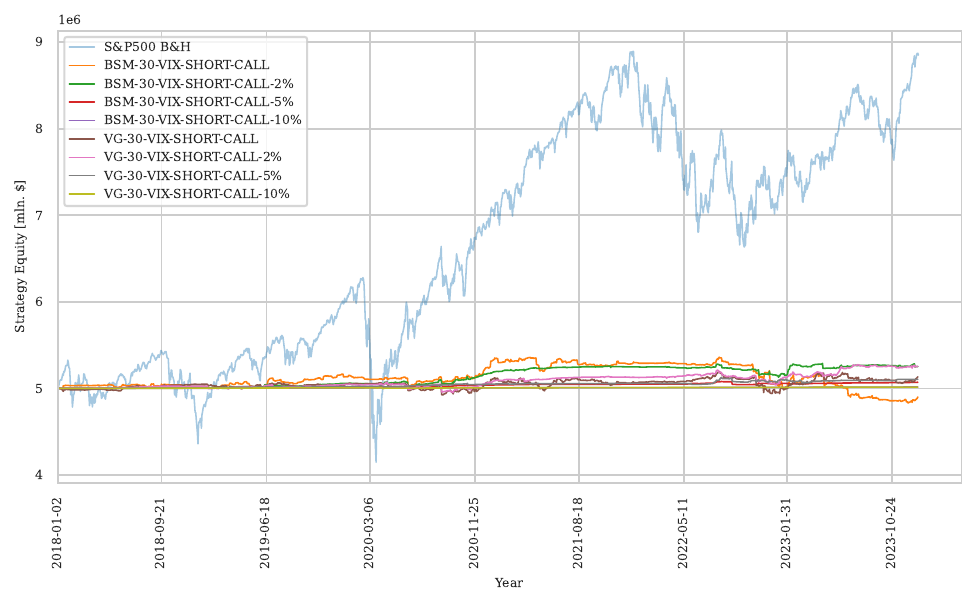} }}
\caption{Intraday Hedged (30 Minutes) Short Call Strategies Equity Lines Comparison}%
\label{fig:30-short-calls}%
\end{figure}

\subsection{Short Puts}
\subsubsection{Hedging Every 130 Minutes}
\begin{figure}[!h]
\centering
\subfloat[\centering Short Puts with Delta Sizing Hedged Every 130 Minutes]{{\includegraphics[width=0.8\columnwidth]{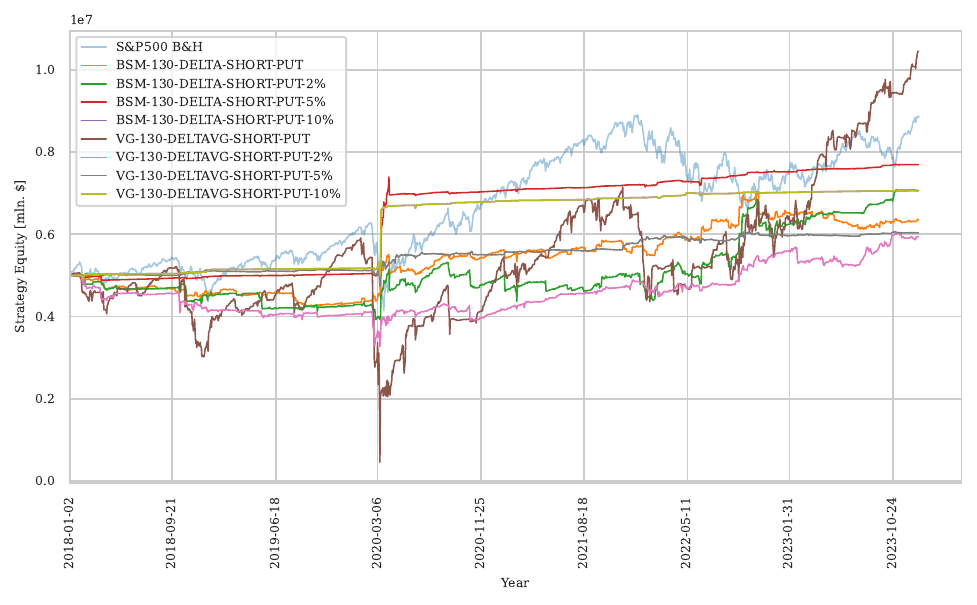} }}\\
\subfloat[\centering Short Puts with VIX Sizing Hedged Every 130 Minutes]{{\includegraphics[width=0.8\columnwidth]{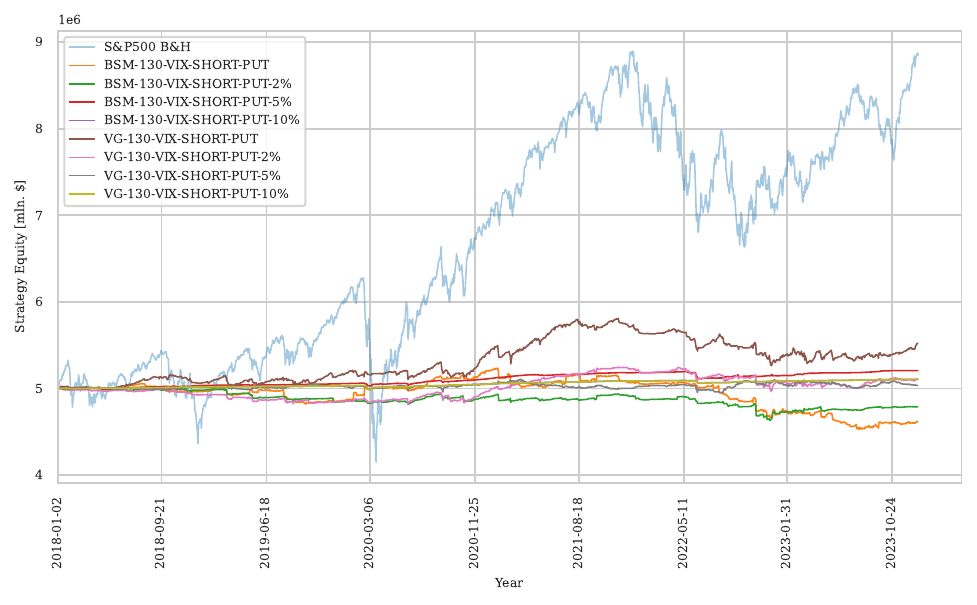} }}
\caption{Intraday Hedged (130 Minutes) Short Put Strategies Equity Lines Comparison}%
\label{fig:130-short-puts}%
\end{figure}

\clearpage
\subsubsection{Hedging Every 30 Minutes}
\begin{figure}[!h]
\centering
\subfloat[\centering Short Puts with Delta Sizing Hedged Every 30 Minutes]{{\includegraphics[width=0.88\columnwidth]{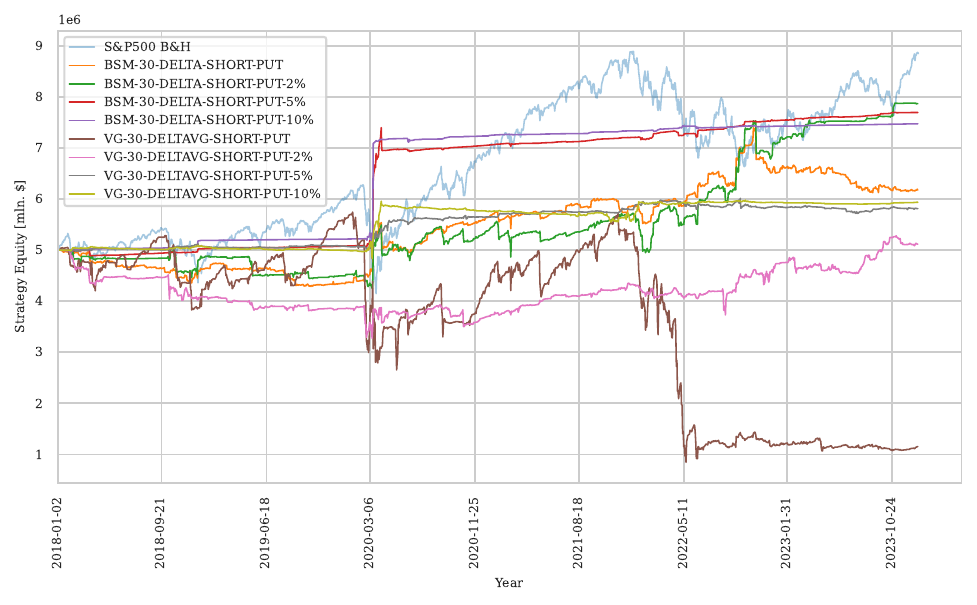} }}\\
\subfloat[\centering Short Puts with VIX Sizing Hedged Every 30 Minutes]{{\includegraphics[width=0.88\columnwidth]{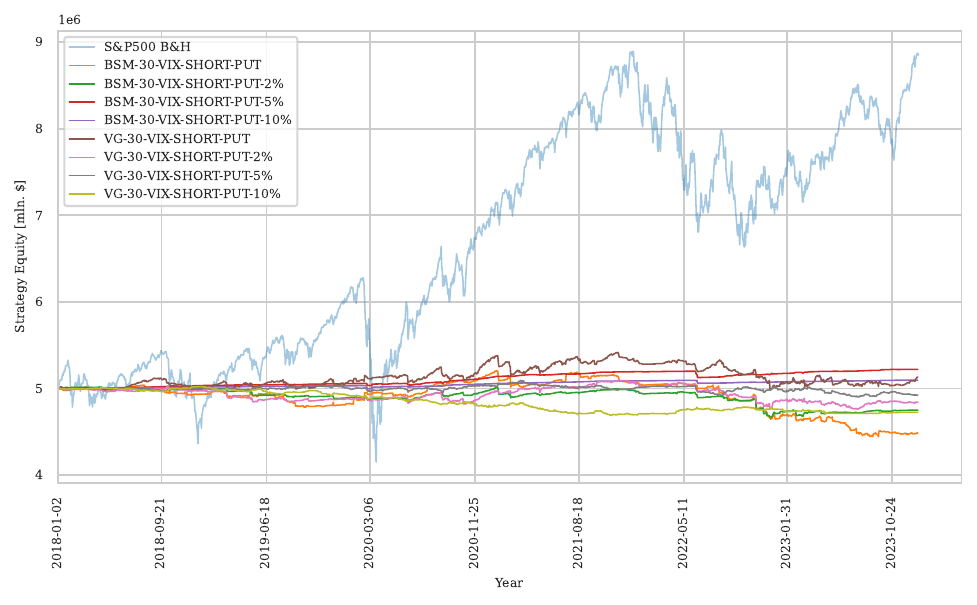} }}
\caption{Intraday Hedged (30 Minutes) Short Put Strategies Equity Lines Comparison}%
\label{fig:30-short-puts}%
\end{figure}

\subsection{Short Strangles}
\subsubsection{Hedging Every 130 Minutes}
\begin{figure}[!h]
\centering
\subfloat[\centering Short Strangles with Delta Sizing Hedged Every 130 Minutes]{{\includegraphics[width=0.8\columnwidth]{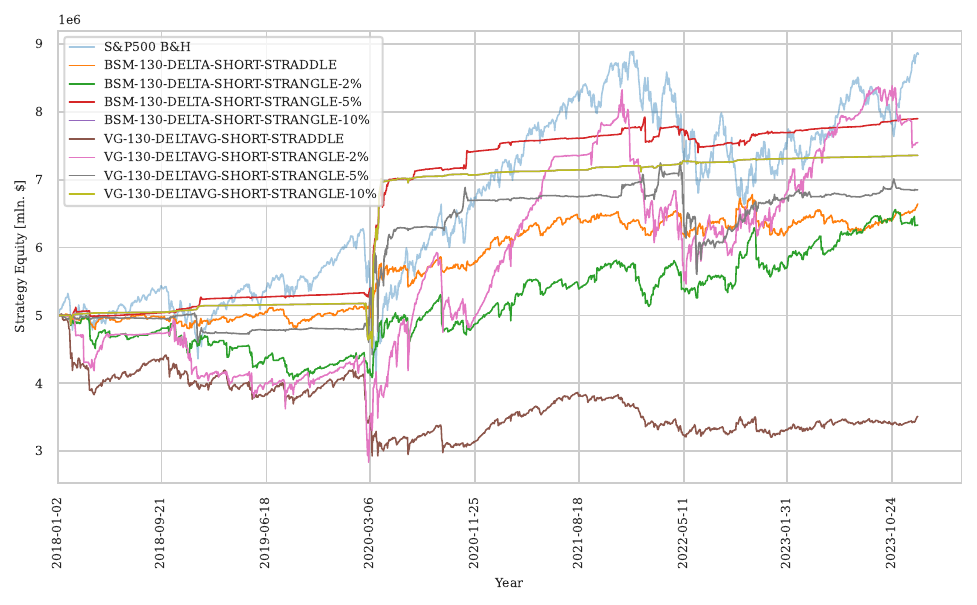} }}\\
\subfloat[\centering Short Strangles with VIX Sizing Hedged Every 130 Minutes]{{\includegraphics[width=0.8\columnwidth]{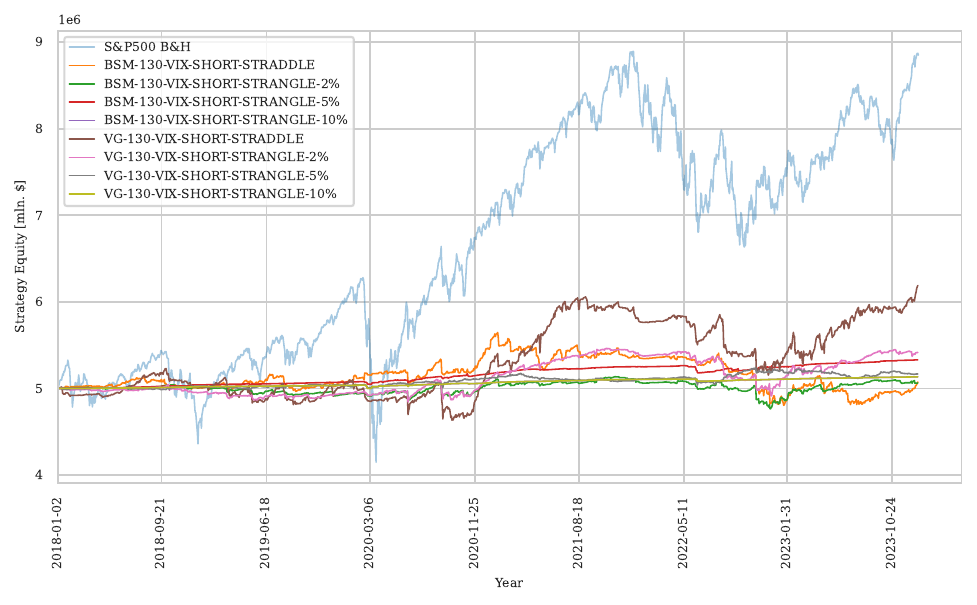} }}
\caption{Intraday Hedged (130 Minutes) Short Strangle Strategies Equity Lines Comparison}%
\label{fig:130-short-strangles}%
\end{figure}

\clearpage
\subsubsection{Hedging Every 30 Minutes}
\begin{figure}[!h]
\centering
\subfloat[\centering Short Strangles with Delta Sizing Hedged Every 30 Minutes]{{\includegraphics[width=0.88\columnwidth]{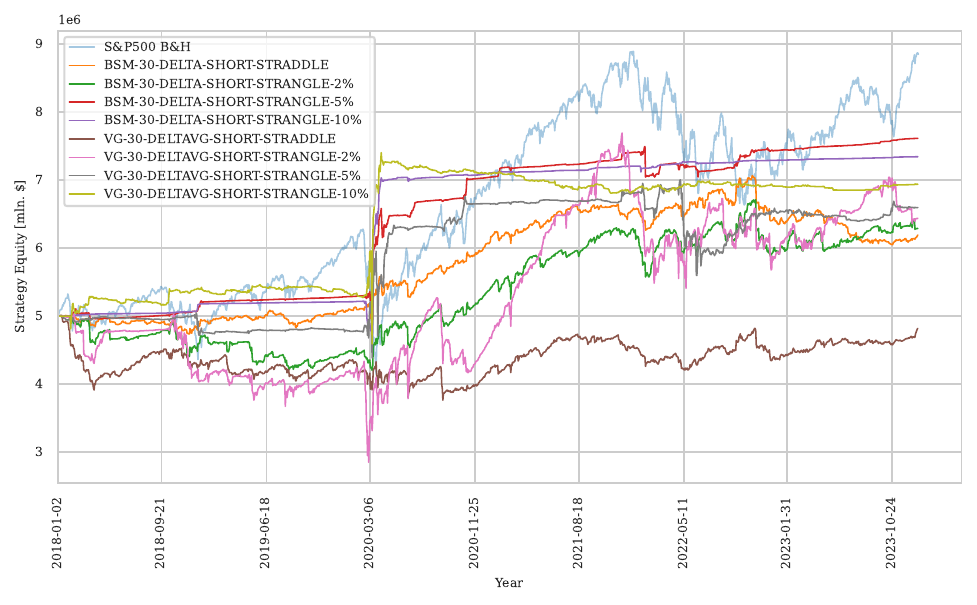} }}\\
\subfloat[\centering Short Strangles with VIX Sizing Hedged Every 30 Minutes]{{\includegraphics[width=0.88\columnwidth]{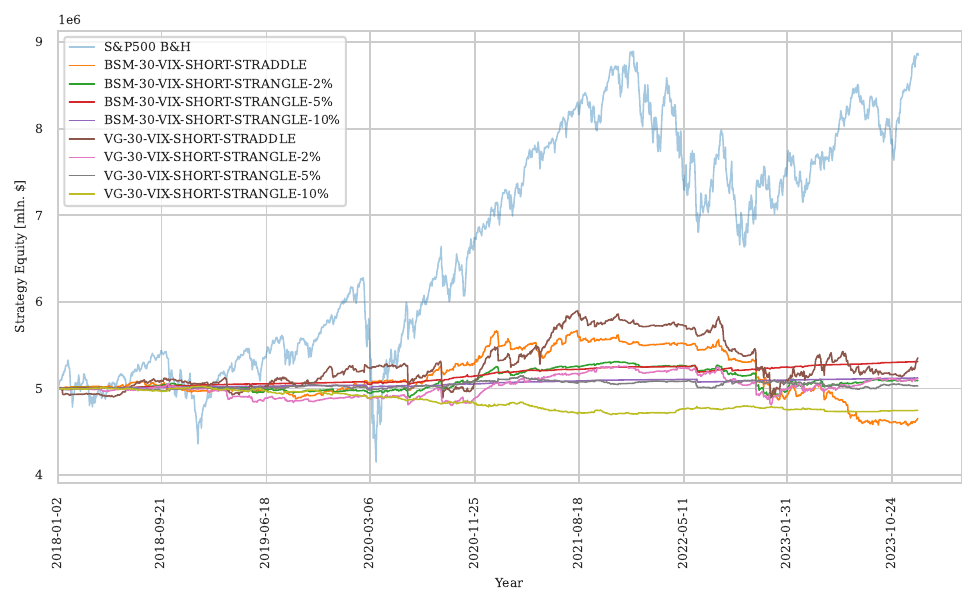} }}
\caption{Intraday Hedged (30 Minutes) Short Strangle Strategies Equity Lines Comparison}%
\label{fig:30-short-strangles}%
\end{figure}

\end{document}